\documentclass[a4paper,11pt]{article}
\usepackage{jcappub} 
\usepackage{lineno}

\newcommand{\lammfp}{\lambda_{\text{mfp}}}
\newcommand{\guvb}{\Gamma_{\text{UVB}}}
\newcommand{\lya}{Ly$\alpha$\ }

\arxivnumber{} 
\title{\boldmath A Robust Estimator for the Lyman-$\alpha$ Forest 1D Power Spectrum at High Redshift}

\author[a,b,c]{M. Wolfson}
\author[d,e]{J.F. Hennawi}
\author[f]{F.B. Davies}
\author[g]{and Z. Luki\'{c}}
\affiliation[a]{Center for Cosmology and AstroParticle Physics, The Ohio State University,\\
191 West Woodruff Avenue, Columbus, OH 43210, USA}
\affiliation[b]{Department of Physics, The Ohio State University,\\
191 West Woodruff Avenue, Columbus, OH 43210, USA}
\affiliation[c]{Department of Astronomy, The Ohio State University,\\
4055 McPherson Laboratory, 140 W 18th Avenue, Columbus, OH 43210, USA}
\affiliation[d]{Department of Physics, University of California,\\
Santa Barbara, CA 93106, USA}
\affiliation[e]{Leiden Observatory, Leiden University,\\
Niels Bohrweg 2, 2333 CA Leiden, Netherlands}
\affiliation[f]{Max-Planck-Institut f\"{u}r Astronomie,\\
K\"{o}nigstuhl 17, 69117 Heidelberg, Germany}
\affiliation[g]{Lawrence Berkeley National Laboratory,\\
1 Cyclotron Rd, Berkeley, CA 94720, USA}

\emailAdd{wolfson.63@osu.edu}

\abstract{
The Lyman-$\alpha$ (Ly$\alpha$) forest flux power spectrum $P(k)$ is sensitive to a range of astrophysical and cosmological effects, including the thermal state of and large-scale fluctuations in the intergalactic medium and the nature of dark matter.
Measuring $P(k)$ becomes increasingly challenging at high redshift, where fewer quasar spectra are available and higher Ly$\alpha$ optical depths result in lower transmitted flux. 
In addition, observational effects and missing data complicate standard Fourier-transform estimators.
Here, we validate a likelihood-based estimator for $P(k)$. 
The method constructs the pixel-level covariance matrix of the transmitted flux field using precomputed $k$-space integrals that account for instrumental resolution and pixelization, enabling direct maximum-likelihood inference of $P(k)$. 
We test the estimator using mock Ly$\alpha$ forest spectra at $z \geq 5$, including analytic Gaussian realizations generated from known power spectra and forward-modeled simulation spectra incorporating observational effects relevant to high-redshift quasar surveys, such as finite spectral resolution, pixel sampling, noise, and missing data.
We compare the recovered power spectra with both the input spectra and standard Fourier-transform estimates.
We find that the likelihood-based estimator recovers the input power spectrum across the tested scales well within expected error and remains robust to the observational effects considered. 
This work establishes a framework for measuring Ly$\alpha$ forest clustering with a robust, nearly optimal power spectrum estimator, with particular focus on quantifying large-scale fluctuations that may arise from ultraviolet background fluctuations or the late stages of reionization.
}

\begin{document}
\maketitle
\flushbottom

\section{Introduction}

The reionization of neutral hydrogen in the intergalactic medium (IGM), driven by the first luminous sources, represents one of the final major phase transitions in cosmic history. 
Despite its importance for the thermal and ionization history of the Universe, the timing, duration, and spatial structure of reionization remain uncertain. 
Measurements of the cosmic microwave background (CMB) constrain the integrated Thomson optical depth, providing an indirect constraint on the midpoint of reionization of $z_{\rm re}=7.7\pm0.7$ \citep{planck_2018}. 
Measurements of the Lyman-$\alpha$ (Ly$\alpha$) forest, which arises from absorption by neutral hydrogen in the spectra of distant quasars \citep{gunn_peterson_1965, lynds_1971}, provide a complementary probe of the IGM \citep[see, e.g.,][for reviews]{meiksin_2009_review, mcquinn_2016_review}.

At $z\gtrsim 5$, measurements of the \lya forest reveal strong sightline-to-sightline variations in transmission and rapid evolution in the effective optical depth, suggesting a late and highly inhomogeneous end to reionization that may extend to $z\sim 5.3$ \citep{becker_2015, becker_2018, eilers_2018, bosman_2018, yang_2020, kashino_2020, bosman_2021_data}. 
Additional constraints from dark pixel statistics \citep{fan_2006, mcgreer_2011, mcgreer_2015, davies_2026} and stacked spectra around \lya and Ly$\beta$ dark gaps \citep{zhu_2024, spina_2024} further support this picture. 
The enhanced scatter in the \lya forest opacity at $z > 5.5$ cannot be explained by density fluctuations alone \citep{bosman_2021_data}, motivating models that invoke spatial fluctuations in the ultraviolet background (UVB) \citep{davies_furlanetto_2016, gnedin_2017, daloisio_2018} and the IGM temperature \citep{daloisio_2018, davies_2018_thermal, wu_2019}. 
These fluctuations can arise naturally during a late, extended, and inhomogeneous reionization process, as ionization fronts heat different regions of the IGM at different times and ionizing photons are absorbed after traveling finite distances. 
Although alternative reionization histories can also explain the observed opacity fluctuations \citep[see, e.g.,][]{asthana_2024}, these mechanisms motivate investigating large-scale fluctuations in the \lya forest as a probe of the ionization and thermal structure of the IGM.

The one-dimensional \lya forest flux power spectrum, $P(k)$, quantifies the amplitude of transmitted flux fluctuations on different spatial scales. 
Measurements of $P(k)$ have been used to constrain a range of physical processes and cosmological parameters, including the thermal state of the IGM, the timing of reionization, the nature of dark matter, and the properties of dark energy \citep[e.g.][]{boera_2019, villasenor_2023, irsic_2024, gonzalez-hernandez_2025, garcia-gallego_2025, garza_2026}. 
The impact of large-scale fluctuations on \lya forest statistics has also been investigated using both the power spectrum and the real-space autocorrelation function, its Fourier-space counterpart \citep{zuo_1992_a, zuo_1992_b, croft_2004, meiksin_2004, mcdonald_2005, gontcho_2014, pontzen_2014_a, pontzen_2014_b, daloisio_2018, meiksin_2019, onorbe_2019, wolfson_2022, gaikwad_2023_mfp, wolfson_2023_xqr30, wolfson_23_thermal, maity_2026, maity_2026_modeling, spina_2026}. 
Robustly measuring $P(k)$ across a range of scales is therefore important for extracting the physical information encoded in the \lya forest.

Accurately measuring $P(k)$ requires a robust method for estimating the power spectrum from observed flux fluctuations. 
The standard approach uses Fourier transforms of the transmitted flux field, while likelihood-based methods estimate $P(k)$ by maximizing the likelihood of the observed flux fluctuations given a model for their covariance.
These methods originate from power spectrum estimation techniques developed for CMB analyses \citep{tegmark_1996, bond_1998} and were first applied to the \lya forest by \cite{mcdonald_2006}. 
Subsequent work developed and applied related likelihood-based approaches to \lya forest power spectrum measurements \citep[see e.g.,][]{palanque-delabrouille_2013, karacayli_2022, karacayli_2025_desi}. 
Although these methods provide an alternative to direct Fourier-transform estimates, their performance depends on the likelihood assumptions, optimization procedure, and treatment of the data. 
Validating these choices is particularly important when measuring power across a broad range of scales and at high redshift, where the \lya forest opacity is high and the transmitted flux field is more strongly non-Gaussian.

Although the \lya forest power spectrum has been measured extensively at lower redshifts \citep[e.g.][]{palanque-delabrouille_2013, irsic_2017_p1d, walther_2018, gaikwad_2020_thermal, karacayli_2022, ravoux_2023}, measurements at $z\gtrsim 5$ remain limited. 
Recent analyses by the Dark Energy Spectroscopic Instrument (DESI) collaboration have also demonstrated the use of multiple methods to measure the power spectrum from a large survey at $z < 5$ \citep{karacayli_2025_desi, ravoux_2025_desi}.
Previous measurements by \cite{viel_2013} reported power spectra at $z=5.0$ and $z=5.4$, while \cite{boera_2019} measured the power spectrum at $z=5.0$.
More recent observations, including the XQR-30 sample \citep{dodorico_2023_xqr30} and the ongoing GHOSTLy observing program \citep{bosman_2026_ghostly}, offer opportunities to extend power spectrum measurements to higher redshifts and improve their precision.
Robust estimation methods are needed to fully exploit these data, particularly in the presence of observational challenges such as missing data, which can complicate measurements on large scales where spatial fluctuations in the IGM may leave a measurable imprint on the power spectrum.

In this work, we validate a likelihood-based method for measuring $P(k)$ across a wide range of scales at high redshift. 
We test the estimator using mock spectra designed to represent different levels of large scale IGM fluctuations.
To model these fluctuations, we use a semi-numerical prescription for UVB fluctuations. 
The UVB fluctuations, characterized by the ionizing photon mean free path, $\lammfp$, imprint themselves on the \lya forest transmitted flux through their effect on the \lya optical depth, $\tau_{\rm Ly\alpha}$, where
\begin{equation}
    \tau_{\rm Ly\alpha} = n_{\rm HI}\sigma_{\rm Ly\alpha} \propto \frac{1}{\Gamma_{\rm HI}} \propto \frac{1}{\lammfp^{\alpha}},
\end{equation}
with $3/2 < \alpha < 2$ \citep[see, e.g.][]{Rauch_1998, haardt_madau_2012}.
By comparing the likelihood-based measurements with the known input power spectrum or measurements obtained using Fourier transforms, we assess the accuracy and robustness of the estimator across different mock datasets and power spectrum bins. 
This validation provides a foundation for applying the method to observational measurements of the high-redshift \lya forest and investigating the impact of large-scale IGM fluctuations on its power spectrum.

The remainder of this paper is organized as follows. 
Section \ref{sec:mock_data} describes the hydrodynamic simulation, the different $\lammfp$ models, the analytic $P(k)$ models, and the construction of realistic mock spectra. 
Section \ref{sec:methods} presents the power spectrum estimators and other analysis methods. 
Section \ref{sec:results} presents the validation results for the different mock datasets. 
Finally, Section \ref{sec:conclusion} summarizes our findings and conclusions.

\section{Mock data}
\label{sec:mock_data}

\subsection{Simulations} 
\label{sec:sim_data}

In this work, we use a hydrodynamical simulation performed with the \texttt{Nyx} code \citep{almgren_2013}, which was specifically developed for modeling the \lya forest and incorporates the updated physical rates described by \cite{lukic_2015}. 
The simulation volume has a comoving box size of $L_{\rm box}=100$ cMpc $h^{-1}$ and contains $4096^3$ dark matter particles and $4096^3$ baryonic grid cells. 
The box is evolved with a spatially uniform UV background from \cite{haardt_madau_2012}, which is switched on at $z \sim 15$.
Simulation outputs are available at redshifts $z=5.0$, 5.5, and 6.0. 
While we focus primarily on results at $z=5.5$, we generate mock datasets spanning $5 \leq z \leq 6$ in steps of $\Delta z = 0.1$, yielding a total of 11 redshift bins. 
For redshifts without a dedicated simulation output, we use the nearest available snapshot and evaluate the proper box size and mean density at the target redshift. 
In other words, the density fluctuations, temperature field, and peculiar velocities are taken directly from the nearest \texttt{Nyx} output, while cosmological scales are computed at the desired redshift.

We additionally use boxes of fluctuating $\guvb$ values generated with the semi-numerical method of \cite{davies_furlanetto_2016}. 
These boxes use the density field from the \texttt{Nyx} simulation at $z = 5.5$ which have a comoving box size of $L_{\rm box}=100$ cMpc $h^{-1}$, but are sampled on a coarser $128^3$ grid. 
We use this one box of $\guvb$ fluctuations at $z = 5.5$ for all redshifts considered in this work.

The method of \cite{davies_furlanetto_2016} produces a spatially varying mean free path by coupling the ionizing background to the local density field according to $\lambda \propto \guvb^{2/3}\Delta^{-1}$, where $\lambda$ is the local mean free path and $\Delta$ is the local matter overdensity. 
The resulting simulations are normalized such that the volume-averaged mean free path satisfies $\langle \lambda \rangle = \lammfp$.
A brief summary of the procedure is as follows.
For this work, the density field is taken from the \texttt{Nyx} simulation snapshot at $z = 5.5$
Dark matter halos are then identified using the method of \cite{mesinger_furlanetto_2007}, assuming a minimum halo mass of $M_{\rm min}=2\times10^9,M_\odot$. 
Galaxies are assigned to halos by abundance matching to the UV luminosity function of \cite{bouwens_2015}. 
The ionizing luminosity of each galaxy is subsequently assumed to be proportional to its UV luminosity, with the proportionality constant treated as a free parameter.
The specific intensity of the ionizing background, $J_\nu$, is then computed using a radiative transfer algorithm, and the photoionization rate, $\guvb$, is obtained by integrating over $J_\nu$. 
Additional details of the method can be found in \cite{davies_furlanetto_2016}, \cite{davies_2018_uvb}, and \cite{davies_2022}.
In this work, we use UVB boxes with $\lammfp =$ 5, 10, 15, 30, 50, and 150 cMpc.

To generate the flux skewers, we combine $\guvb$ skewers extracted from the UVB boxes with \texttt{Nyx} skewers drawn from the corresponding locations in the hydrodynamical simulation. 
Because the UVB boxes have a lower spatial resolution than the \texttt{Nyx} simulation, we linearly interpolate $\log(\guvb)$ onto the \texttt{Nyx} skewer grid before combining the two skewers.
We generate 1000 skewers at fixed locations within the comoving box, using the same skewer locations for each combination of $\lammfp$ and $z$.

\begin{figure}
    \centering
	\includegraphics[width=.9\textwidth]{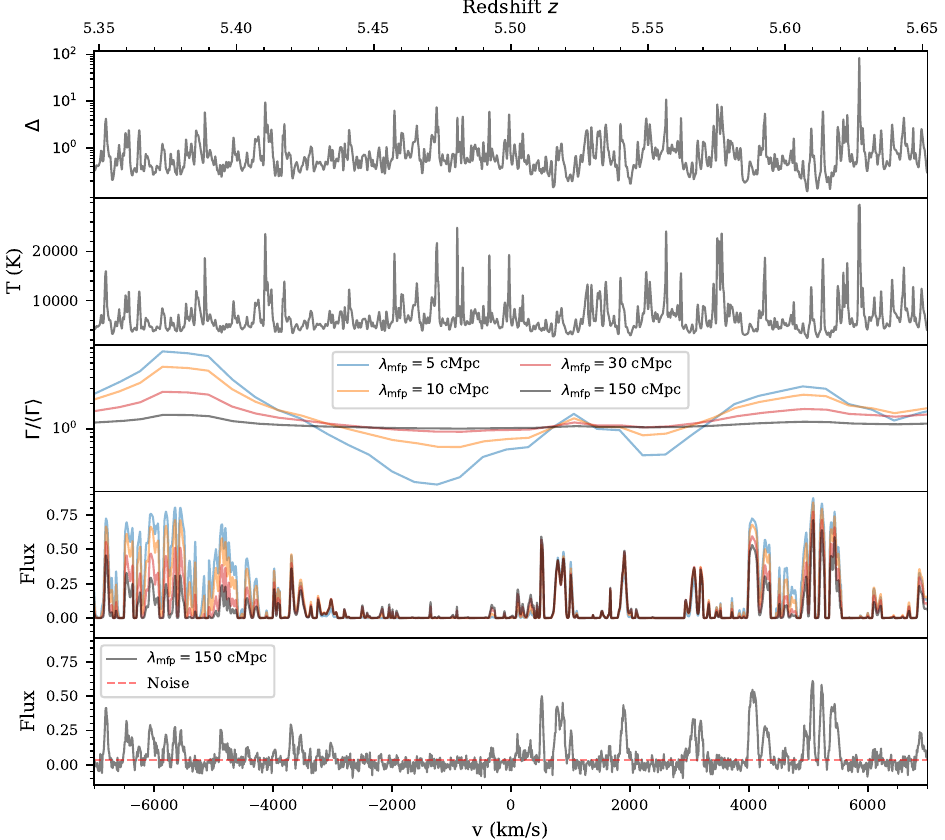}
    \caption{
        Example sightline through the simulation showing the impact of UVB fluctuations on the \lya forest flux.
        From top to bottom, the panels show: 
            (1) the matter overdensity, $\Delta$; 
            (2) the gas temperature, T; 
            (3) the normalized photoionization rate, $\Gamma/\langle\Gamma\rangle$, for four values of the mean free path, $\lammfp$, 5, 10, 30, and 150 cMpc; 
            (4) the corresponding transmitted flux fraction along the same sightline; and 
            (5) the transmitted flux for the $\lammfp = 150$ cMpc model after forward modeling to a spectral resolution of $R=8800$ and a signal-to-noise ratio of $\mathrm{SNR}=30$. Also included in this panel is a dashed red line showing the $1\sigma$ noise level, $\sigma_{F} \simeq 0.033$ (see Equation \eqref{eq:snr_f}).
    }
    \label{fig:skewer_fields}
\end{figure}

Figure \ref{fig:skewer_fields} shows an example sightline from the simulation at $z = 5.5$. 
The first three panels display the underlying simulation fields: the matter overdensity, $\Delta$, the gas temperature, $T$, and the normalized photoionization rate, $\Gamma/\langle\Gamma\rangle$. 
For visual clarity, we show only four of the six mean free path models considered in this work, corresponding to $\lammfp = 5$, 10, 30, and 150 cMpc. 
The omitted $\lammfp = 15$ and 50 cMpc models exhibit fluctuations that lie between the neighboring cases shown.
The fourth panel illustrates the resulting transmitted \lya forest flux along the same sightline. 
Regions with enhanced photoionization rates correspond to reduced neutral hydrogen fractions and therefore increased transmission.
This relationship can be seen, for example, near $v \simeq -5000$ km s$^{-1}$, where a positive fluctuation in $\Gamma/\langle\Gamma\rangle$ coincides with a region of enhanced transmitted flux.
The dependence on $\lammfp$ is also apparent in this region, with the smaller $\lammfp$ models exhibiting substantially larger fluctuations in both $\Gamma/\langle\Gamma\rangle$ and the transmitted flux.
All flux skewers have been normalized to reproduce the same mean transmitted flux, $\langle F \rangle$, such that the differences shown arise solely from fluctuations in the UVB.
The bottom panel shows the $\lammfp = 150$ cMpc transmission skewer after forward modeling to include instrumental resolution and noise, as described in Section \ref{sec:forward_modeling}

\begin{figure}
    \centering
	\includegraphics[width=\textwidth]{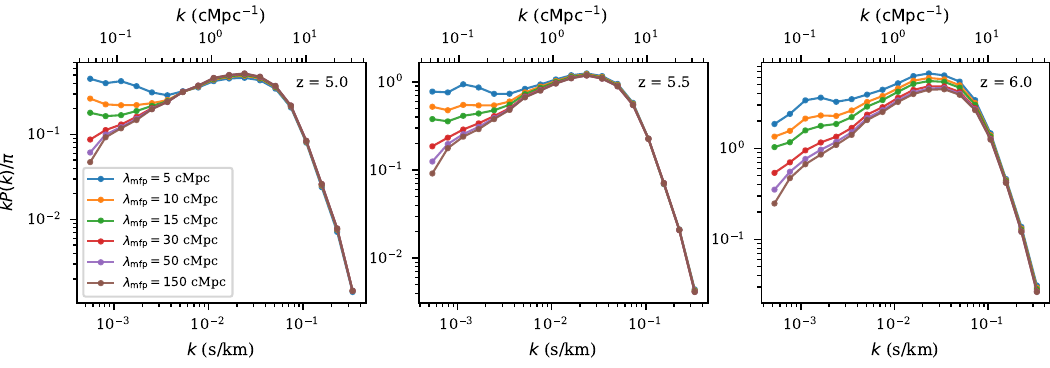}
    \caption{
        Dimensionless \lya forest flux power spectrum, $kP(k)/\pi$, at $z = 5.0$, $5.5$, and $6.0$.
        The power spectra are computed by averaging the FFT estimates from 1,000 simulated skewers without forward modeling, binned into 18 logarithmically spaced $k$ bins.
        Shorter $\lammfp$ models exhibit enhanced large-scale power (small-$k$) as a result of stronger UVB fluctuations.
    }
    \label{fig:power_models_bins}
\end{figure}

To illustrate the impact of $\lammfp$ on the \lya forest flux power spectrum, Figure \ref{fig:power_models_bins} shows the power spectra measured from simulated skewers using a Fourier transform approach at $z = 5.0$, $5.5$, and $6.0$. 
Each curve is obtained by averaging the power spectra of 1,000 skewers, binned into 18 logarithmically spaced $k$ bins, for each of the six $\lammfp$ values considered in this work. 
As expected, models with shorter $\lammfp$ exhibit enhanced large-scale (small $k$) power due to stronger UVB fluctuations. 
The scales over which this enhancement is significant depend on redshift:
at $z = 5.0$, the enhancement is apparent at $k \lesssim 5 \times 10^{-3}$ s/km; 
at $z = 5.5$, it extends to $k \lesssim 10^{-2}$ s/km; 
and at $z = 6.0$, it is evident across a broader range of scales, up to $k \lesssim 10^{-1}$ s/km. 
This redshift dependence reflects the changing relative contributions of UVB fluctuations and density fluctuations to the flux power spectrum. 
At lower redshifts, the growth of structure increases the contribution of density fluctuations, such that UVB fluctuations dominate primarily on larger scales.

Because the UVB fluctuations are introduced through post-processing, the underlying density, temperature, and velocity fields are identical across all models. 
Consequently, the power spectra converge on small scales (large $k$), with negligible differences at $k \gtrsim 10^{-1}$ s/km. 
However, these models are sufficient for the goals of this work, which focuses on validating the method for measuring the \lya forest power spectrum from data, particularly in the presence of enhanced large-scale power due to UVB fluctuations.

\subsection{Analytical Mocks}
\label{sec:analytic_mocks}

In addition to the simulation-based mocks described in Section \ref{sec:sim_data}, we generate Gaussian random field realizations from a known input power spectrum. 
These mocks provide a controlled test case in which the true underlying power spectrum is known exactly, allowing us to compare the recovered power spectrum to the input analytic form and quantify the performance of our analysis methods.

For the power spectrum, $P(k)$, we assume a power law with an exponential cutoff:
\begin{equation}
P_{\mathrm{1D}}(k)
=
A
\left(\frac{k}{k_0}\right)^n
\exp\left[
-\left(\frac{k}{k_s}\right)^\alpha
\right]
\label{eq:five_param_power}
\end{equation}
where $A$ sets the overall normalization, $k_0$ is a fixed reference scale, $n$ controls the large-scale spectral slope, $k_s$ is the characteristic cutoff scale associated with the suppression of small-scale power, and $\alpha$ determines the sharpness of the high-$k$ cutoff.
A power law was chosen for its simplicity, while the exponential cutoff was motivated by the suppression of power observed at high \(k\) \citep[see, e.g.,][]{viel_2013, irsic_2017_wdm, walther_2018, boera_2019}.

\begin{figure}
    \centering
	\includegraphics[width=.5\textwidth]{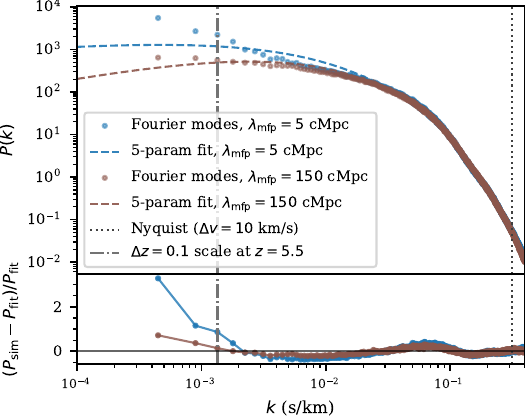}
    \caption{
        Mean unbinned \lya forest flux power spectra, $P(k)$, measured from the $z = 5.5$ simulation mocks for $\lammfp = 5$ and 150 cMpc (blue and brown points, respectively).
        The dashed curves show the corresponding best-fit analytic models defined by Equation \eqref{eq:five_param_power}.
        The bottom panel shows the fractional residuals, $(P_{\rm raw} - P_{\rm fit})/P_{\rm fit}$, between the simulation $P(k)$ and their best-fit models.
        The dotted vertical line shows the Nyquist frequency corresponding to a spectrograph with $\Delta v = 10$ s/km pixels, while the dot-dashed vertical line marks the scale corresponding to $\Delta z = 0.1$ at $z=5.5$. 
        These scales are relevant for the mock datasets considered in this work and are discussed further in Section \ref{sec:forward_modeling}.
    }
        \label{fig:analytic_fit}
    \end{figure}

To obtain realistic parameter values, we fit Equation \eqref{eq:five_param_power} to the mean un-binned power spectrum measured from the $z = 5.5$ simulation mocks for $\lammfp = 5$ and 150 cMpc. 
These two cases span the range of large-scale power observed in our simulation suite and therefore provide representative models for testing our methodology. 
The best-fit parameters are listed in Table \ref{tab:p1d_fit_params}, and the corresponding fits are shown in Figure \ref{fig:analytic_fit}.

\begin{table*}[ht]
    \centering
    \label{tab:p1d_fit_params}    
    \begin{tabular}{|lcccccc|}
        \hline
        Model &
        $A$ &
        $k_0\ [{\rm s\,km}^{-1}]$ &
        $n$ &
        $k_s\ [{\rm s\,km}^{-1}]$ &
        $\alpha$ &
        ${\rm RMSE}_{\log_{10}}$
        \\
        \hline

        $\lambda_{\rm mfp}=5\,{\rm Mpc}$ &
        3835.5 &
        0.00722 &
        0.232 &
        0.00216 &
        0.5 &
        0.0661
        \\

        $\lambda_{\rm mfp}=150\,{\rm Mpc}$ &
        5965.1 &
        0.0218 &
        0.584 &
        0.00178 &
        0.5 &
        0.0444
        \\

        \hline
        {Lower bounds} &
        $1\times10^{-12}$ &
        $4.48\times10^{-4}$ &
        $-6.0$ &
        $4.48\times10^{-4}$ &
        $0.5$ &
        --
        \\

        {Upper bounds} &
        $1\times10^{8}$ &
        $4\times10^{-1}$ &
        $2.0$ &
        $4\times10^{-1}$ &
        $6.0$ &
        --
        \\

        \hline
    \end{tabular}
    \caption{
        Best-fit parameters for the analytic \lya forest flux power spectrum model defined by Equation \eqref{eq:five_param_power}.
        The model is fit to the mean unbinned power spectra of the simulation mocks for $\lammfp = 5$ and 150 cMpc at $z = 5.5$.
        The bottom section of the table shows the lower and upper bounds of the model parameters imposed when fitting the data. 
    }
\end{table*}

The fits to Equation \eqref{eq:five_param_power} are performed using \texttt{scipy.optimize.curve\_fit}, which implements a non-linear least-squares optimization. 
We fit the model in logarithmic space by minimizing the difference between $\log_{10} P(k)$ from the simulation and the corresponding model prediction, using data restricted to $k < 4 \times 10^{-1}$ s/km (the range shown in Figure \ref{fig:analytic_fit}).
Parameter bounds used in the optimization are listed in Table \ref{tab:p1d_fit_params}. 
These bounds are chosen to be broad while remaining physically reasonable. 
The lower bounds on $k_0$ and $k_s$ correspond approximately to the smallest accessible $k$ mode in the simulation volume. 
The upper bound is set just beyond the Nyquist frequency associated with $\Delta v = 10$ km s$^{-1}$ sampling, which matches the spectrograph resolution adopted in the forward-modeling procedure described in Section \ref{sec:forward_modeling}.
None of the fitted parameters approach their imposed bounds, with the exception of $\alpha$, which saturates the lower limit of 0.5 in both cases. 
This indicates that the data prefer a relatively shallow high-$k$ cutoff within the flexibility allowed by the model. 
However, since the resulting fits reproduce the simulated power spectra well over the scales of interest (as shown in the bottom panel of Figure \ref{fig:analytic_fit}), and our analysis focuses primarily on large-scale power differences, we do not expand the parameter bounds or modify the functional form.
Because the fitting is performed in log-space, the procedure effectively weights fractional deviations approximately uniformly across the dynamic range of the power spectrum, which spans several orders of magnitude in amplitude.

A goodness-of-fit metric is reported in the last column of Table \ref{tab:p1d_fit_params}, defined as the root-mean-square difference between the simulated and fitted power spectra in logarithmic space:
\begin{equation}
    \text{RMSE}_{\log_{10}} = \sqrt{\frac{1}{N} \sum_i \left( \log_{10}P_{\text{sim}}(k_i) - \log_{10}P_{\text{fit}}(k_i)\right)^2}
    \label{eq:rmse_log}
\end{equation}

For the $\lammfp = 5$ cMpc model, we get $\mathrm{RMSE}_{\log_{10}} = 0.0611$, corresponding to a typical multiplicative deviation of $10^{0.0611} \approx 1.15$, or $\sim 15\%$ in $P(k)$. 
For the $\lammfp = 150$ cMpc model, we get $\mathrm{RMSE}_{\log_{10}} = 0.0444$, corresponding to $10^{0.0444} \approx 1.11$, or $\sim 11\%$ deviations in $P(k)$.
These values indicate that Equation \eqref{eq:five_param_power} provides a good phenomenological description of the simulated power spectra over the fitted $k$-range, particularly given that the models are intended to span the range of large-scale $P(k)$ behavior produced by the UVB-fluctuation simulations.
As shown in Figure \ref{fig:analytic_fit}, the largest residuals occur at the lowest $k$ modes, where the simulation exhibits a sharper rise in power. 
These scales are also the most affected by sample variance due to the finite number of unbinned modes in the simulation volume. 
As is expected from the simulation skewers, the fit to $\lammfp = 5$ cMpc has greater large-scale (small $k$) power than the fit to the $\lammfp = 150$ cMpc, so we will be able to explore two analytic models with different behaviors of large-scale power. 

These analytic fits provide a sufficiently realistic and smooth representation of the \lya forest power spectrum, while offering greater flexibility for controlled tests than the simulation models alone. 
They enable direct comparisons between different power-spectrum estimation methods, which may be sensitive to slightly different scales within a given $k$ bin, as discussed further in Section \ref{sec:proj_power}. 
Furthermore, because the analytic functions can be evaluated at arbitrarily large scales, they allow us to test the sensitivity of our measurements to power on scales beyond the range of discrete Fourier modes available in the simulations.

The next step is to generate $\delta F$ skewers from the analytic power spectrum. 
To do so, we construct covariance matrices corresponding to the fit $P(k)$ models. 
The velocity-space two-point correlation function, $\xi(v)$, is obtained by Fourier transforming  $P(k)$:
\begin{equation}
    \xi(v) = \frac{1}{\pi} \int_0^\infty P(k)\cos(kv)\,dk,
    \label{eq:analytic_autocorr}
\end{equation}
which is evaluated numerically using adaptive quadrature. 
To accurately capture the scale dependence of the correlation function, $\xi(v)$ is computed on a finely sampled grid of velocity separations $v$ spanning the range of interest.

The covariance matrix between pixels in a mock skewer, $C_{ij}$, is then given by
\begin{equation}
C_{ij} = \xi(|v_i - v_j|),
\label{eq:analytic_cov}
\end{equation}
where $v_i = i\,\Delta v$ defines the pixel grid for a skewer of length $L$ with pixel spacing $\Delta v$. 
In practice, we first evaluate $\xi(v)$ on a uniform grid and then use cubic interpolation to obtain values at the required pairwise separations.
Gaussian realizations of the field are then generated by drawing samples from a multivariate normal distribution with zero mean and covariance matrix $C_{ij}$ using \texttt{numpy.random.multivariate\_normal}. 
This procedure directly produces realizations of the flux contrast field, $\delta_F = \frac{F-\langle F\rangle}{\langle F\rangle}$, rather than separate density or flux fields requiring additional transformations or normalization by $\langle F \rangle$.

We want each sightline mock skewer generated from this analytic procedure to represent the information in a typical quasar sightline at $z \geq 5.0$ with $\Delta z = 0.1$.
At $z = 5.0$ using $\Delta = 0.1$ gives a total length per sightline of $L = 5059$ km/s while at $z = 6.0$ using $\delta z = 0.1$ gives a total length per sightline of $L = 4329$ km/s.
For simplicity here we generate mock spectra with total length $L = 4000$ km/s and a pixel size of $\Delta v$ = 10 km/s, matching the expected pixel spacing for XQR-30 data \citep{dodorico_2023_xqr30}.

\begin{figure}
    \centering
	\includegraphics[width=.5\textwidth]{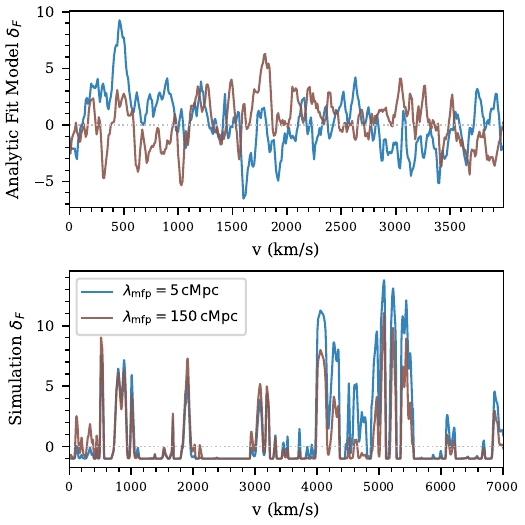}
    \caption{
        Top panel: Example flux-contrast skewers, $\delta_F$, generated from the analytic power spectrum models of Equation \eqref{eq:five_param_power} using the procedure described in Section \ref{sec:analytic_mocks}.
        Bottom panel: $\delta_F$ skewers from the simulation descibed in Section \ref{sec:sim_data}, shown for the same sightline as Figure \ref{fig:skewer_fields} (restricted to $v \geq 0$ km/s).
        In both panels, the $\lammfp = 5$ and 150 cMpc models are shown in blue and brown, respectively. 
        The shorter mean free path model exhibits larger-amplitude fluctuations on large scales, consistent with the enhanced large-scale power seen in Figure \ref{fig:analytic_fit}.
        Because the analytic mocks are generated as Gaussian realizations from the input power spectrum, they are not constrained by the physical bounds of the transmitted flux ($0 \leq F \leq 1$) and can therefore produce $\delta_F$ values outside the range expected for physical \lya forest spectra. 
    }
    \label{fig:analytic_df}
\end{figure}

Figure \ref{fig:analytic_df} shows example $\delta_F$ skewers generated from the analytic power spectrum models alongside a skewer from the simulations.
The analytic realizations are shown in the top panel, while the bottom panel shows the same simulation sightline presented in Figure \ref{fig:skewer_fields} (restricted to $v \geq 0$ km/s). 
In both panels, the $\lammfp = 5$ and 150 cMpc models are shown in blue and brown, respectively. 
Although the analytic mocks are not expected to reproduce the detailed structure of any individual simulation sightline, they capture the qualitative dependence of the flux fluctuations on $\lammfp$. 
In particular, the $\lammfp = 5$ cMpc model exhibits larger-amplitude fluctuations on large scales, consistent with the enhanced large-scale power seen in Figure \ref{fig:analytic_fit}. 
This behavior is the primary feature that the analytic mocks are designed to reproduce.

Because the analytic mocks are generated as Gaussian realizations of the input power spectrum in Equation \eqref{eq:five_param_power}, they do not capture the non-Gaussianity of the transmitted flux field and are not constrained by the physical bounds on the transmitted flux ($0 \leq F \leq 1$), allowing them to produce $\delta_F < -1$.
However, these differences do not affect the suitability of these Gaussian analytic mocks for validating the power-spectrum estimation methods considered in this work, as the goal is to compare the recovered power spectrum with the input spectrum, which here is known exactly by construction.

\subsection{Forward-Modeling}
\label{sec:forward_modeling}

Depending on the analysis being performed, we consider three different versions of forward modeling.

For the analytic mocks described in Section \ref{sec:analytic_mocks}, the primary goal is to compare power-spectrum estimation methods against a known analytic input $P(k)$ model. 
In this case, realistic instrumental effects are unnecessary and would only complicate the interpretation of the results. 
However, a small diagonal contribution to the covariance matrix improves numerical stability when performing matrix operations. 
We therefore add Gaussian white noise to the $\delta_F$ field with a standard deviation of $\sigma_{\delta F} = 0.2$, which is equivalent to adding a small diagonal noise term to the covariance matrix.
For a \lya forest at $z = 5.5$ with $\langle F\rangle=0.0591$, the equivalent signal-to-noise ratio per $\Delta v = 10$ km s$^{-1}$ pixel ($\text{SNR}_{\Delta v}$) for this would be $\sim 85$ (see Equation \eqref{eq:snr_df}). 

Beyond the analytic mocks, we also consider simulation-based mocks intended to mimic realistic observations. 
In particular, we are interested in datasets with properties similar to the XQR-30 survey \citep{dodorico_2023_xqr30}. 
We therefore construct two sets of forward-modeled simulation mocks: a simple model with fixed resolution and signal-to-noise ratio, and a more realistic model that reproduces the resolution and noise properties of the individual XQR-30 sightlines. 
Both sets of mocks are generated from the simulation skewers described in Section \ref{sec:sim_data}.

For the simple mocks, we adopt a spectral resolution of $R = 8800$ and a $\text{SNR}_{\Delta v}$ of 30 at all redshifts. 
The instrumental resolution is modeled by smoothing the flux skewers with a Gaussian kernel corresponding to $\mathrm{FWHM}\approx34$ km s$^{-1}$. 
The smoothed spectra are then resampled onto a grid with $\Delta v = 10$ km s$^{-1}$ pixels.
For simplicity, we assume flux-independent Gaussian noise. 
We generate a single noise realization consisting of 1000 skewers, each with values drawn from a Gaussian distribution with standard deviation
\begin{equation}
    \sigma_F = \frac{1}{\mathrm{SNR}_{\Delta v}}.
    \label{eq:snr_f}
\end{equation}
This same noise realization is applied to all models and redshifts so that differences between the resulting power spectra arise solely from the underlying signal rather than from variations in the noise realization. 
For the flux-contrast field the corresponding noise level is
\begin{equation}
    \sigma_{\delta_F}
    =
    \frac{1}{\mathrm{SNR}_{\Delta v}\langle F\rangle},
    \label{eq:snr_df}
\end{equation}
which evaluates to $\sigma_{\delta_F}=0.564$ for $\langle F\rangle=0.0591$ at $z=5.5$.

Finally, we construct mocks that more closely reproduce the XQR-30 observations. 
At each redshift, we use the measured spectral resolutions and noise vectors associated with the observed sightlines. 
The measured resolutions are generally larger than the nominal $R=8800$ resolution of the X-Shooter VIS arm and are listed in Table 3 of \cite{dodorico_2023_xqr30}. 
For each simulation skewer, we generate a separate realization corresponding to every available quasar sightline. 
The flux is first smoothed to the appropriate resolution for that sightline. 
Noise is then generated from a Gaussian distribution with unit variance and scaled by the observed per-pixel uncertainty vector, thereby reproducing the wavelength-dependent noise properties of the corresponding XQR-30 spectrum.
These mocks also use the exact masking of pixels present in the data.
Because the observational noise properties and masking are imposed independently of the simulated flux field, these mocks do not capture correlations between noise levels or masked regions and the underlying \lya absorption, such as the association between the masking of damped \lya systems (DLAs) and enhanced absorption in the real data.
Nevertheless, they provide an appropriate test of the effects of realistic noise, resolution, and masking on the $P(k)$ measurement, which is the primary goal of this analysis.

To ensure that our validation tests represent a realistic amount of information from the XQR-30 dataset, we adopt mock dataset sizes that approximately match the observational sample. 
For the analytic and simple simulation mocks, we use approximately 30 sightlines worth of information without explicitly reproducing the redshift-dependent number of sightlines in the data. 
For the analytic mocks, this corresponds to 30 skewers of length $L=4000$ km/s per mock dataset. 
For the simple simulation mocks, the full simulation box has a skewer length of $L_{\rm box}=100$ cMpc $h^{-1}$, which is longer than a single sightline spanning $\Delta z=0.1$ over the redshift range considered here. 
At $z=5.5$, for example, a $\Delta z=0.1$ interval corresponds to approximately $32$ cMpc $h^{-1}$. 
We retain the full periodic length of the simulation skewers to avoid introducing non-periodic boundary effects into the FFT-based estimates. 
We therefore use 10 full-length skewers for these mocks, providing a comparable total path length to approximately 30 shorter sightlines while preserving the periodic boundary conditions. 
Although the distribution of information across scales differs between these cases, this provides a sufficient approximation for the purposes of this validation.
For the XQR-30-like mocks, we instead use the noise vectors and masks from the observational data directly, so that the number and length of sightlines exactly match the data in each redshift bin. 
The number of sightlines increases from 16 at $z=5$ to 33 at $z=5.6$, before decreasing to 12 at $z=6$.

\section{Methods}
\label{sec:methods} 

\subsection{Fourier Transform}
\label{sec:fourier_transform}

The one-dimensional \lya forest power spectrum, $P(k)$, is defined by the Fourier transform of $\delta_F$. 
In continuous form, the power spectrum is defined as
\begin{equation}
P(k) = \left\langle \left| \tilde{\delta}_F(k) \right|^2 \right\rangle,
\label{eq:power_ft}
\end{equation}
where $\tilde{\delta}_F(k)$ is the Fourier transform of $\delta_F(v)$ and the angle brackets denote an ensemble average over sightlines.
In practice, the measured $P(k)$ is affected by both the finite instrumental resolution and the pixel size of the data. 
These effects are described by the window function, $W_R(k)$,
\begin{equation}
    W_R (k, \sigma_R, \Delta v) = 
    \exp\!\left[-\frac{1}{2}(k \sigma_R)^2 \right]
    \frac{\sin(k \Delta v / 2)}{(k \Delta v / 2)} .
    \label{eq:window_function}
\end{equation}
where the first term describes smoothing by the Gaussian line-spread function and the second term accounts for the finite pixel width $\Delta v$. 
Both effects suppress power on small scales (large $k$).

In addition to instrumental effects, the observed power spectrum contains a contribution from noise. 
For uncorrelated Gaussian white noise with standard deviation $\sigma_{\delta F}$, the corresponding noise power is independent of $k$ and is given by
\begin{equation}
    P_{\mathrm{noise}}(k) = \Delta v \sigma_{\delta F}^{2},
    \label{eq:noise_power}
\end{equation}
where $\Delta v$ is the pixel spacing in velocity units.
Combining these effects, the corrected estimate of the underlying $P(k)$ is
\begin{equation}
    P(k) = 
    \left\langle 
        \frac{P_{\mathrm{raw}}(k) - P_{\mathrm{noise}}(k)}
             {W_R^2(k, \sigma_R, \Delta v)}
    \right\rangle ,
    \label{eq:power_correction}
\end{equation}
where the averaging is taken over independent sightlines or spectra within a redshift bin.
This correction subtracts the white-noise contribution and deconvolves the effects of instrumental resolution and pixel size through $W_R$, resulting in an estimate of the intrinsic $P(k)$.

In practice, the Fourier transform is computed numerically using a Fast Fourier Transform (FFT), which efficiently evaluates the discrete Fourier modes of a skewer sampled on a regular velocity grid. 
FFT-based estimators provide a simple and computationally efficient measurement of $P(k)$ and are widely used in \lya forest analyses.

However, FFT-based approaches assume data sampled on a uniformly spaced grid with no missing pixels. 
In observational datasets, it is common to mask portions of a spectrum affected by contaminants such as high-column-density damped \lya absorbers, strong metal absorption systems, sky-line residuals, or other data-quality issues. 
These masked regions break the regular sampling required by a standard FFT and can introduce mode mixing and biases in the measured power spectrum if not treated carefully. 
As a result, alternative estimators that can naturally accommodate irregular sampling or missing data may be advantageous for analyses involving heavily masked spectra.
\cite{boera_2019}, for example, used a Lomb-Scargle algorithm \citep{lomb_1976, scargle_1982} to make their $P(k)$ measurement which is an extension of Fourier analysis for irregularly sampled data for their measurement at $z = 5$. 
Here we instead investigate a maximum likelihood approach.

\subsection{Likelihood Optimization}
\label{sec:likelihood_method}

As an alternative to FFT-based $P(k)$ estimators, likelihood-based approaches have been used to measure the \lya forest power spectrum \citep{mcdonald_2006, palanque-delabrouille_2013, karacayli_2022, karacayli_2025_desi}. 
The method implemented in this work follows \cite{palanque-delabrouille_2013} and is based on likelihood techniques originally developed for CMB power spectrum estimation \citep{tegmark_1996, bond_1998}, which have also been applied to weak lensing \citep{seljak_1998} and the \lya forest \citep{mcdonald_2006}.
The method estimates the power spectrum by directly maximizing the likelihood of the observed flux fluctuations as a function of the bandpowers.
 
For this implementation, we consider the normalized flux transmission fraction in pixel $i$ to be
\begin{equation}
    {\delta_F}_i = \frac{F_i}{\langle F \rangle_i} - 1 = s_i + n_i,
    \label{eq:delta_f_signal_noise}
\end{equation}
where $s_i$ represents the signal and $n_i$ the noise contribution.  
We assume that the signal and noise are independent, each with zero mean and covariance matrices defined by:
\begin{align}
    C^{\mathrm{S}}_{ij} &= \langle s_i s_j \rangle, \\
    C^{\mathrm{N}}_{ij} &= \langle n_i n_j \rangle = \sigma_i \sigma_j \, \delta_{ij},
\end{align}
where $\delta_{ij}$ is the Kronecker delta and $\sigma_i$ is the observed noise at pixel $i$.
The total covariance matrix is therefore
\begin{equation}
    C_{ij} = \langle {\delta_F}_i {\delta_F}_j \rangle = \langle s_i s_j + s_i n_j + s_j n_i + n_i n_j \rangle = \langle s_i s_j \rangle + \langle n_i n_j \rangle = C^{\mathrm{S}}_{ij} + C^{\mathrm{N}}_{ij}.
    \label{eq:covar_def_sum}
\end{equation}

The signal covariance $C^{\mathrm{S}}_{ij}$ describes the correlation between transmitted \lya forest flux fluctuations in pixels $i$ and $j$, separated by a velocity lag $\Delta v (i-j)$.
Assuming statistical homogeneity along the line of sight, this covariance depends only on the pixel separation and is therefore equivalent to the flux autocorrelation function. 
The autocorrelation function and the one-dimensional flux power spectrum are Fourier transforms pf each other, allowing the signal covariance to be expressed in terms of the power spectrum as follows:
\begin{equation}
    C^{\mathrm{S}}_{ij}
    = \int_{-\infty}^{+\infty} P_{\mathrm{1D}}(k)\, e^{-ik \Delta v (i-j)}\, \mathrm{d}k 
    = \int_{0}^{+\infty} 2\, P_{\mathrm{1D}}(k)\, \cos[k \Delta v (i-j)]\, \mathrm{d}k.
\end{equation}

In practice, we evaluate the power spectrum in a discrete set of $N_\ell$ wavenumber bins $(k_{\ell-1}, k_\ell)$ and approximate $P_{\mathrm{1D}}(k)$ as constant within each bin, with the value of the $\ell$th band power denoted by
\[
P_\ell \equiv P\left(k_{\ell \text{effective}}\right).
\]
With this discretization, the signal covariance can be written as a finite sum:
\begin{equation}
    C^{\mathrm{S}}_{ij}(P) = \sum_{\ell=1}^{N_\ell} 
    P_\ell \int_{k_{\ell-1}}^{k_\ell} 
    2\, \cos[k \Delta v (i-j)]\, \mathrm{d}k.
    \label{eq:cov_discrete}
\end{equation}

Similar to the discussion in Section \ref{sec:fourier_transform}, observed data has a suppression in small scale $P(k)$ from spectrograph resolution and pixel size. 
Therefore we also have to include the window function, $W_R$, defined in Equation \eqref{eq:window_function}, in this integral as well:
\begin{equation}
    C^{\mathrm{S}}_{ij}(P) = 
    \sum_{\ell=1}^{N_\ell} P_\ell 
    \int_{k_{\ell-1}}^{k_\ell} 2 \cos[k \Delta v (i-j)] 
    W_R(k, R_i, \Delta v) \times W_R(k, R_j, \Delta v)\, \mathrm{d}k.
    \label{eq:cov_discrete_res}
\end{equation}

For a given sightline, this integral describes the contribution of the Fourier modes within each $k$ bin to the covariance between pairs of pixels. 
We refer to this integral as the kernel for that bin. 
The kernels, $K_\ell$, depend only on the sightline's pixel separations and the $k$-bin boundaries, so they remain fixed throughout the optimization. 
They are defined as follows:
\begin{equation}
    (K_{\ell})_{ij} =  \int_{k_{\ell-1}}^{k_\ell} 
    2\, \cos[k \Delta v (i-j)]\, W(k, R_i, \Delta v) W(k, R_j, \Delta v)\, \mathrm{d}k.
    \label{eq:integrals}
\end{equation}
So we can rewrite the covariance matrix as
\begin{equation}
    C^{\mathrm{S}}_{ij}(P) = 
    \sum_{\ell=1}^{N_\ell} P_\ell 
    (K_{\ell})_{ij}.
    \label{eq:cov_sum}
\end{equation}

For a single spectrum $sp$ containing $N_{\mathrm{pix}}^{sp}$ pixels, the likelihood of observing transmitted flux, $\delta_F$, is
\begin{equation}
    \mathcal{L}_{sp}(P) =
    \frac{1}{(2\pi)^{N_{\mathrm{pix}}^{sp}/2} \sqrt{\det C}} 
    \exp\!\left[ -\frac{1}{2} \delta_F^{\mathsf{T}} C^{-1} \delta_F \right].
    \label{eq:df_like}
\end{equation}
Rather than fitting each spectrum individually, we combine all $N_{\mathrm{sp}}$ spectra in a given redshift bin into a joint likelihood,
\begin{equation}
    \mathcal{L}(P(k)) = \prod_{sp=1}^{N_{\mathrm{sp}}} \mathcal{L}_{sp}(P(k)),
    \label{eq:total_like}
\end{equation}
and determine the best-fit power spectrum $P_{opt}(k)$ by maximizing this combined likelihood.

The likelihood in Equation \eqref{eq:df_like} assumes that the transmitted flux fluctuations follow a Gaussian process, whereas the \lya forest is intrinsically non-Gaussian, especially at high-$z$. 
Nevertheless, this Gaussian likelihood can still provide a reliable estimate of $P(k)$. 
In particular, the non-Gaussianity of the flux field does not necessarily introduce bias into the estimated $P(k)$, just as the standard sample variance estimator, $\frac{1}{N-1}\sum_i (x_i-\bar{x})^2$, is unbiased for a non-Gaussian distribution. 
Under appropriate assumptions, maximum-likelihood estimation provides an optimal estimate of the power spectrum for a Gaussian field. 
We therefore expect this approach to remain nearly optimal for the \lya forest and test its performance using analytic mocks and simulations in Section \ref{sec:results}.

\begin{figure}
    \centering
	\includegraphics[width=.5\textwidth]{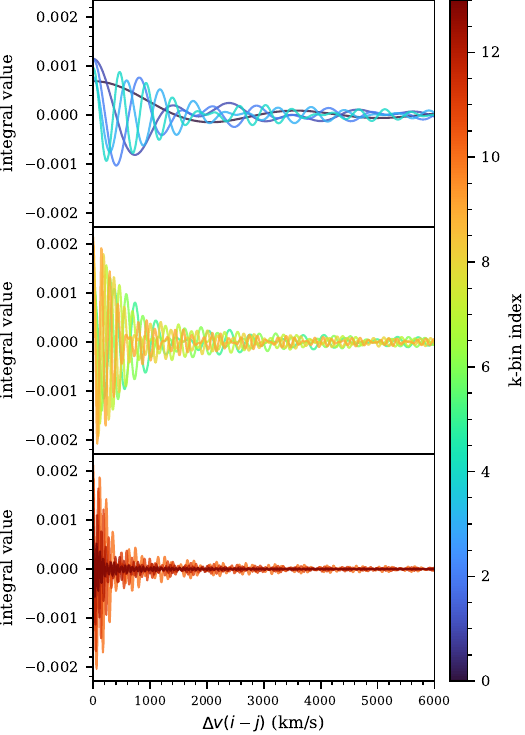}
    \caption{
        Precomputed kernels from Equation \eqref{eq:integrals} as a function of pixel separation, $\Delta v (i-j)$, for the $k$ bins listed in Table \ref{tab:kbins}. 
        These were computed for the constant-SNR forward-modeling setup described in Section \ref{sec:forward_modeling}, with $R=8800$ and $\Delta v = 10$ km/s.
        Each colored curve corresponds to a different $k$-bin, progressing from the lowest $k$ (largest-scale) bin to the highest $k$ (smallest-scale) bin.
        The kernels exhibit oscillatory behavior whose characteristic scale depends on the $k$ values in the bin, while the overall envelope of these oscillations decreases with increasing $\Delta v(i - j)$. 
        During the optimization, these kernels act as a basis to build the autocorrelation of the transmitted flux when weighted by the band power, $P_\ell$ (see Equation \eqref{eq:cov_sum}).
    }
    \label{fig:integrals}
\end{figure}

\begin{table}
    \centering
    \caption{
        Velocity-space $k$ bin edges used for the $P(k)$ measurements. 
        All values are in units of s/km.
    }
    \label{tab:kbins}
    \begin{tabular}{ccc}
        \hline
        Bin & $k_{\rm min}$ & $k_{\rm max}$ \\
        \hline
        0  & $1.00\times10^{-12}$ & $2.1716\times10^{-3}$ \\
        1  & $2.1716\times10^{-3}$ & $5.7910\times10^{-3}$ \\
        2  & $5.7910\times10^{-3}$ & $9.4103\times10^{-3}$ \\
        3  & $9.4103\times10^{-3}$ & $1.1898\times10^{-2}$ \\
        4  & $1.1898\times10^{-2}$ & $1.5042\times10^{-2}$ \\
        5  & $1.5042\times10^{-2}$ & $1.9019\times10^{-2}$ \\
        6  & $1.9019\times10^{-2}$ & $2.4046\times10^{-2}$ \\
        7  & $2.4046\times10^{-2}$ & $3.0401\times10^{-2}$ \\
        8  & $3.0401\times10^{-2}$ & $3.8437\times10^{-2}$ \\
        9 & $3.8437\times10^{-2}$ & $4.8597\times10^{-2}$ \\
        10 & $4.8597\times10^{-2}$ & $6.1442\times10^{-2}$ \\
        11 & $6.1442\times10^{-2}$ & $7.7683\times10^{-2}$ \\
        12 & $7.7683\times10^{-2}$ & $9.8216\times10^{-2}$ \\
        13 & $9.8216\times10^{-2}$ & $1.2418\times10^{-1}$ \\
        14 & $1.2418\times10^{-1}$ & $1.5700\times10^{-1}$ \\
        15 & $1.5700\times10^{-1}$ & $1.9850\times10^{-1}$ \\
        16 & $1.9850\times10^{-1}$ & $2.5097\times10^{-1}$ \\
        17 & $2.5097\times10^{-1}$ & $3.1730\times10^{-1}$ \\
        \hline
    \end{tabular}
\end{table}

To build physical intuition for the covariance construction, first consider the kernels from Equation \eqref{eq:integrals}. 
To illustrate these, we evaluate these kernels for the simple forward-modeling described in Section \ref{sec:forward_modeling}, assuming a constant spectral resolution of $R=8800$.
The kernels depend on the choice of $k$-bins, and the effects of different bin choices on the resulting optimized $P(k)$ values are explored in Appendix \ref{appendix:bins}.
In particular, these tests demonstrate that the lowest-$k$ bin should begin sufficiently below the fundamental mode while extending beyond it, such that the bin contains information on the largest scales accessible to the spectra.
For the main analysis, we additionally require each $k$ bin to contain at least one discrete Fourier mode so that the optimized measurements can be directly compared with the FFT-based measurements.
For the data we want to considered here, with $z = 5.5$ and $\Delta z = 0.1$, we get an expected length of $L = 4666$ km/s and so the fundamental mode is 
\begin{equation}
    k_{\rm min} = \frac{2\pi}{L}
    = \frac{2\pi}{4666~\mathrm{km~s}^{-1}}
    = 1.35 \times 10^{-3}~\mathrm{s\ km}^{-1}.
\end{equation}
The $k$-bin edges used throughout the main analysis are listed in Table \ref{tab:kbins}.

The resulting integrals are shown in Figure \ref{fig:integrals}, where the colors denote the $k$-bin index, with index 0 corresponding to the lowest-$k$ (largest-scale) bin. 
Each integral oscillates with increasing pixel separation, $\Delta v (i-j)$, while the envelope of the oscillations decays with increasing separation. 
The characteristic oscillation scale is set by the $k$ values in the bin considered.
As expected, the lowest-$k$ bin varies most slowly with separation, reflecting the largest spatial scales.

The covariance matrix in Equation \eqref{eq:cov_discrete_res} is constructed by summing the precomputed integrals, each weighted by its corresponding band power, $P_\ell$.
Since the covariance between two pixels depends only on their separation, $\Delta v = |v_i-v_j|$, the covariance matrix can equivalently be interpreted as the auto-correlation function of the $\delta_F$ field evaluated at each pixel separation.
Examples of these auto-correlation functions are shown in the top-right panels of Figures \ref{fig:analytic_one_opt} and \ref{fig:one_opt}, which show results for an analytic mock and a simple forward-modeled mock, respectively.

\subsection{Effective Wavenumber}
\label{sec:eff_k}

The likelihood-based estimator does not measure the power spectrum at a single wavenumber, but instead constrains each bandpower, $P_\ell$, through its contribution to the covariance matrix $C$ in Equation \eqref{eq:df_like}. 
As shown in Equation \eqref{eq:cov_sum}, this contribution is described by the covariance kernel $K_\ell$, defined in Equation \eqref{eq:integrals} as an integral over the wavenumber range of the corresponding bin. 
The sensitivity to different wavenumbers within a bin is not necessarily uniform, as it depends on the covariance weighting, instrumental resolution, and pixel window functions. 
Consequently, the center of a bin does not necessarily correspond to the wavenumber that best represents the measurement.

To characterize the scales probed by each bin, we define an effective wavenumber, $k_{\mathrm{eff},\ell}$, as a Fisher-weighted effective wavenumber. 
The Fisher matrix quantifies the information available about the $P_{\ell}$ through the response of the covariance matrix to changes in their amplitudes \citep[see, e.g.,][]{tegmark_1996, bond_1998}.
We construct the effective wavenumber using the same covariance weighting as the Fisher matrix, with one covariance kernel replaced by its wavenumber-weighted counterpart. 
The resulting effective wavenumber provides a representative scale for each bin and facilitates comparisons with the underlying power spectrum and measurements obtained using the FFT method.

The effective wavenumber, $k_{\rm eff}$, is defined as
\begin{equation}
    k_{\mathrm{eff},\ell}
    =
    \frac{
    \mathrm{Tr}\!\left[
    C_{\mathrm{fid}}^{-1}
    K_{\ell}
    C_{\mathrm{fid}}^{-1}
    K_{\ell}^{k}
    \right]
    }{
    \mathrm{Tr}\!\left[
    C_{\mathrm{fid}}^{-1}
    K_{\ell}
    C_{\mathrm{fid}}^{-1}
    K_{\ell}
    \right]
    }
    \label{eq:k_eff}
\end{equation}
where $K_{\ell}^{k}$ is the wavenumber-weighted covariance kernel, defined as
\begin{equation}
    (K_{\ell}^{k})_{ij}
    =
    \int_{k_{\ell-1}}^{k_\ell}
    k \,
    2 \cos[k \Delta v (i-j)]
    \, W(k,R_i,\Delta v)\,
    W(k,R_j,\Delta v)\,
    \mathrm{d}k,
    \label{eq:k_cov_k_weight}
\end{equation}

This weighting is evaluated using a fiducial covariance,
\begin{equation}
    C_{\mathrm{fid}}
    =
    C^{\mathrm{N}}
    +
    \sum_{\ell}
    P^{\mathrm{fid}}_\ell
    K_{\ell},
    \label{eq:fid_cov}
\end{equation}
where $C^{\mathrm{N}}$ is the noise covariance and $P^{\mathrm{fid}}_\ell$ is the fiducial power in bin $\ell$. 
A fiducial covariance is required because the Fisher weighting depends on the assumed signal and noise covariance. 
The choice of fiducial power spectrum entering $C_{\mathrm{fid}}$ for this computation will be discussed in the relevant sections below.

\subsection{Effective Bandpower}
\label{sec:proj_power}

Because the underlying power spectrum can vary across a bin, evaluating $P(k)$ at a single representative wavenumber does not necessarily capture the power to which the optimization is most sensitive. 
We therefore define a projected effective bandpower, $P_{\mathrm{eff},\ell}$, using the same Fisher-weighting as for $k_{\mathrm{eff},\ell}$. 
This provides a representative power that accounts for the covariance weighting of the estimator when comparing the optimized power with the underlying power spectrum.

The projected effective power is defined as, $P_{\mathrm{eff},\ell}$ as
\begin{equation}
    P_{\mathrm{eff},\ell}
    =
    \frac{
    \mathrm{Tr}\!\left[
    C_{\mathrm{fid}}^{-1}
    K_{\ell}
    C_{\mathrm{fid}}^{-1}
    K_{\ell}^{P}
    \right]
    }{
    \mathrm{Tr}\!\left[
    C_{\mathrm{fid}}^{-1}
    K_{\ell}
    C_{\mathrm{fid}}^{-1}
    K_{\ell}
    \right]
    }.
    \label{eq:P_eff}
\end{equation}
where $C_{\mathrm{fid}}$ is the fiducial covariance defined in Equation \eqref{eq:fid_cov}, and $K_{\ell}^{P}$ is the power-spectrum-weighted covariance kernel,
\begin{equation}
    (K_{\ell}^{P})_{ij}    =
    \int_{k_{\ell-1}}^{k_\ell}
    P(k)\,
    2 \cos[k \Delta v (i-j)]
    \, W(k,R_i,\Delta v)\,
    \times W(k,R_j,\Delta v)\,
    \mathrm{d}k.
    \label{eq:k_cov_Pk_weight}
\end{equation}
This construction applies the same Fisher weighting to the underlying power spectrum, consistently accounting for the effects of noise, instrumental resolution, and covariance weighting across the bin. 
In general, $P_{\mathrm{eff},\ell}$ may not be equal to $P(k_{\mathrm{eff},\ell})$, since the effective bandpower and effective wavenumber are defined by distinct Fisher-weighted averages over the bin. 
The choices of $P(k)$ used to calculate $K_{\ell}^{P}$ and of the fiducial power spectrum entering $C_{\mathrm{fid}}$ for this computation will be discussed in the relevant sections below.

\subsection{Implementation Details}
\label{sec:power_est_implementation}

As we have established the mathematical framework of the $P(k)$ estimator, several implementation choices are important for making the optimization computationally efficient. 
Here we summarize the numerical methods used throughout this work.

The covariance kernels, $K_l$, defined in Equation \eqref{eq:integrals}, along with the weighted kernels, $K_l^k$ and $K_l^P$, defined in Equations \eqref{eq:k_cov_k_weight} and \eqref{eq:k_cov_Pk_weight}, depend only on the velocity grid, spectral resolution, and chosen $k$-bins. 
These matrices are therefore precomputed once for each sightline before the optimization and reused throughout the likelihood evaluation, substantially reducing the computational cost.
This is sightline specific because the resolution in quasar samples is typically sightline specific, even when re-binned to a fixed velocity grid. 

For the Gaussian likelihood in Equation \eqref{eq:df_like}, we use the implementation provided by \texttt{jax.scipy.stats.multivariate\_normal}.
Rather than optimizing the physical band powers directly, we optimize band powers normalized by a fiducial power spectrum and rescale the result after convergence. 
We use a fiducial power spectrum that is the $\lammfp = 150$ cMpc simulation model value for the $z$ considered, though we confirm that using a different $\lammfp$ value does not impact the results. 

We optimize the posterior probability and adopt a uniform prior on the band powers. 
The dimensionless band powers, $kP(k)/\pi$, are restricted by $\vert kP(k) / \pi \vert < 10^9$. 
This bound is intentionally conservative and does not affect the recovered solutions; it is included only to avoid numerical instabilities associated with extremely large covariance matrix elements.
In addition, we require the signal contribution to the covariance matrix diagonal ($i=j$) to remain positive for every sightline, excluding unphysical parameter combinations with negative variance.

The posterior is maximized by minimizing the negative log-posterior using the Adam gradient-based optimizer \citep{kingma_2017} as implemented in \texttt{optax.adam}. 
Gradients with respect to the normalized band powers are computed using \texttt{JAX} automatic differentiation. 
The optimization begins from the fiducial power spectrum and proceeds for a user-specified number of iterations.
In practice, we use 1000 optimization steps for the analytic mocks and 1200 steps for the simple simulation and XQR-30 forward-modeled mocks.
At each iteration, the negative log-posterior and its gradient are evaluated, the optimizer updates the normalized band powers, and the parameter vector yielding the lowest negative log-posterior encountered during the optimization is retained as the final solution.

\section{Results}
\label{sec:results}

We now compare measurements of $P(k)$ obtained with the FFT method described in Section \ref{sec:fourier_transform} and the likelihood optimization method described in Section \ref{sec:likelihood_method}. 
We perform this comparison using three classes of mock data: the analytic $P(k)$ mocks from Section \ref{sec:analytic_mocks}, the simple forward-modeled simulation mocks, and the XQR-30-like forward-modeled simulation mocks.
Both of these latter two mocks come from the simulation described in Section \ref{sec:sim_data}.
All forward-modeling is described in Section \ref{sec:forward_modeling}.

\subsection{Analytic Power Mocks}
\label{sec:analytic_mock_results}

We first consider the mocks from the analytic models described in Section \ref{sec:analytic_mocks} for both $\lammfp = 5$ cMpc and $\lammfp = 150$ cMpc. 
We generate 100 independent mock datasets, each consisting of 30 skewers to represent a 30 quasar sightlines with $\Delta z = 0.1$ as is expected from an XQR-30 like dataset, and measure the power spectrum of each dataset using both the likelihood optimization and FFT-based estimators.

For each mock dataset, we obtain an optimized measurement (including the noise term in Equation \eqref{eq:covar_def_sum}), $P_{\mathrm{opt}}(k)$, and a noise-corrected FFT measurement, $P_{\mathrm{FFT}}(k)$. 
Note that these mocks are not forward-modeled for noise and resolution, so we do not correct for a window function $W_R$ in any of the calculations discussed in this section. 
Because the mocks are generated from a known analytic power spectrum, we can compare both measurements directly to the input $P(k)$ to assess how accurately each method recovers the underlying spectrum. 
However, this comparison requires different definitions of the reference power in each $k$ bin to account for differences in how the two methods weight scales within each bin.

For the FFT method, we evaluate the analytic $P(k)$ at the discrete Fourier modes available for each skewer and average these values within each $k$ bin, following the same procedure used to bin the mock power spectra. 
This approach accounts for the finite set of Fourier modes contributing to each bin. 
The resulting reference values can differ by a few percent from $P(k_{\mathrm{FFT,eff}})$, where $k_{\mathrm{FFT,eff}}$ is the mean of the discrete Fourier modes in the bin, particularly at high $k$.

As discussed in Section \ref{sec:proj_power}, the optimization method weights scales within each $k$ bin according to the kernels defined in Equation \eqref{eq:integrals}. 
Therefore, the appropriate reference power for this method is the effective power, $P_{\mathrm{eff}}$, defined in Equation \eqref{eq:P_eff}. 
Computing this quantity requires evaluating the kernels weighted by the analytic power spectrum, as defined in Equation \eqref{eq:k_cov_Pk_weight}, using the analytic model in Equation \eqref{eq:five_param_power}.
The calculation also requires a fiducial covariance, which depends on a fiducial power spectrum through Equation \eqref{eq:cov_sum}. 
We define this fiducial power spectrum to be the analytic model evaluated at the effective wavenumber obtained from Equation \eqref{eq:k_eff}.
The effective wavenumber itself is also computed using a fiducial covariance and power spectrum; for this calculation, we evaluate the analytic power spectrum at the arithmetic mean of the $k$-bin boundaries. 
We verify that modest variations in the fiducial power spectrum used to construct these covariances do not significantly affect the resulting $k_{\mathrm{eff}}$ or $P_{\mathrm{eff}}$.
The resulting $P_{\mathrm{eff}}$ values can differ by up to $\sim 10\%$ from $P(k_{\mathrm{eff}})$ at both the smallest and largest $k$ bins. 

\begin{figure*}
    \centering
	\includegraphics[width=\textwidth]{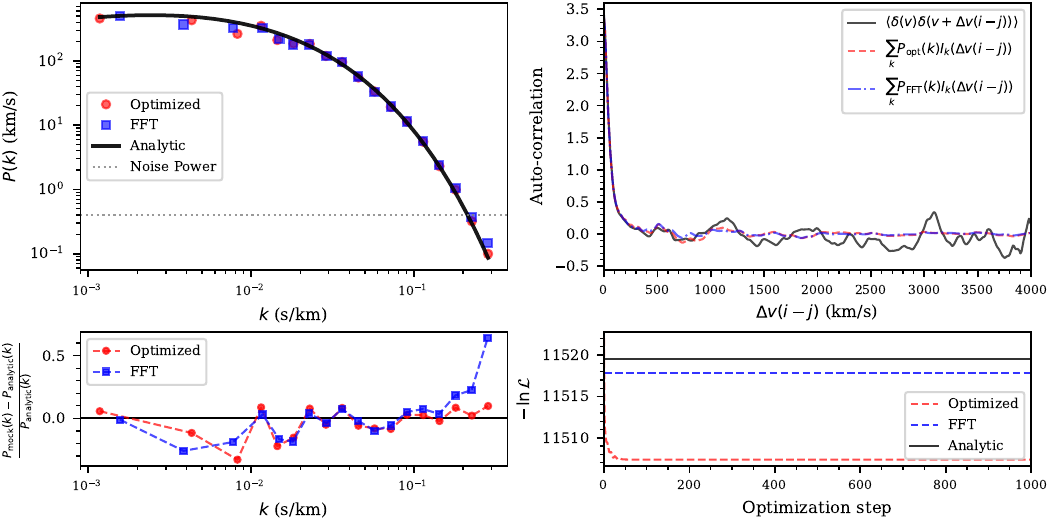}
    \caption{
        Validation on a mock dataset from the $\lammfp = 150$ cMpc analytic model comparing FFT and optimization-based measurements of $P(k)$.
        Top left: Measured $P(k)$ from the FFT method corrected for noise, $P_{\rm FFT}(k)$, (blue squares) and the optimization method, $P_{\rm opt}(k)$, (red circles). 
        In addition there is the analytic function as a black line and the contribution from the Gaussian noise power as a dashed grey horizontal line. 
        Bottom left: Fractional difference in $P(k)$ relative to the analytic function for FFT (blue) and optimization (red). 
        Top right: Autocorrelation function from the mock skewers (black), compared to reconstructions from Eq. \eqref{eq:cov_sum} using $P_{\rm opt}(k)$ (red) and $P_{\rm FFT}(k)$ (blue). 
        Bottom right: Minimization of the negative log-likelihood from Equation \eqref{eq:total_like} during optimization (red dashed), compared to the likelihood evaluated at $P_{\rm FFT}(k)$ (blue dashed), and the analytic value $P_{\rm eff}$ (black solid).
    }
    \label{fig:analytic_one_opt}
\end{figure*}

An example optimization result is shown in Figure \ref{fig:analytic_one_opt} for one $\lammfp = 150$ cMpc analytic mock dataset. 
The left panels show the power spectrum measurements obtained using the FFT method described in Section \ref{sec:fourier_transform} (blue squares) and the optimization method described in Section \ref{sec:likelihood_method} (red circles) (top), along with their residuals relative to the corresponding reference model values defined earlier in this section (bottom). 
The top-left panel also shows the amplitude of the white-noise term added to the skewers for numerical stability (grey dotted line), which becomes significant in the highest-$k$ bins.

The top right panel shows the autocorrelation function of the mock dataset (black solid line) and the autocorrelation functions calculated from the signal covariance matrix in Equation \eqref{eq:cov_sum}, using $P_{\mathrm{opt}}(k)$ and $P_{\mathrm{FFT}}(k)$ (red and blue dashed lines, respectively). 
The comparison demonstrates that the covariance matrix defined in Equation \eqref{eq:cov_sum} provides a reasonable description of the autocorrelation function, particularly at small pixel separations.

Finally, the bottom right panel shows the negative log-posterior as a function of optimization step (red dashed line). 
The horizontal blue dashed and black solid lines show the negative log-posterior evaluated using $P_{\mathrm{FFT}}(k)$ and $P_{\mathrm{eff}}$, respectively. 
The optimized solution achieves the lowest negative log-posterior among the three cases, as expected for the maximum-posterior solution.

\begin{figure*}
    \centering
	\includegraphics[width=\textwidth]{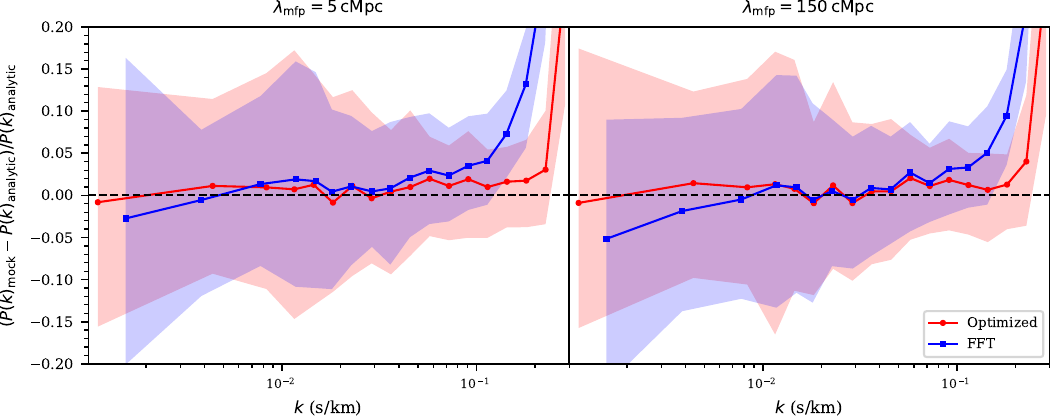}
    \caption{
        The distribution of the difference between the mock $P(k)$ and the analytic $P(k)$ function value for both the optimization method (red) and the FFT method (blue). 
        The left column shows mock results from the analytic model for $\lammfp = 5$ cMpc while the right column shows mock results from the analytic model for $\lammfp = 150$ cMpc. 
        Both panels show a solid line with the mean value over 100 mock datasets while the shaded region denotes the 16th to the 84th percentile for the $1 \sigma$ region. 
    }
    \label{fig:analytic_opt}
\end{figure*}

We next examine the distribution of residuals across 100 mock datasets to assess whether either method exhibits systematic deviations from the analytic power spectrum. 
Figure \ref{fig:analytic_opt} shows the residuals for the noise-corrected FFT method (blue squares) and the optimization method (red circles), relative to the analytic reference power defined above for each method. 
The solid lines indicate the mean residual across the mocks, while the shaded regions show the 16th to 84th percentile range. 
Results for $\lammfp = 5$ cMpc and $\lammfp = 150$ cMpc are shown in the left and right panels, respectively.

The scatter in the residuals reflects both variation among Gaussian realizations of the underlying power spectrum and any measurement uncertainty associated with each method. 
For both $\lammfp$ models, the two methods exhibit comparable scatter and residuals consistent with zero for $k < 1 \times 10^{-1}$ s/km, indicating good recovery of the underlying analytic power spectrum on these scales. 
For the optimization method, the residual remains consistent with zero to $k \lesssim 2 \times 10^{-1}$ s/km, with a significant deviation appearing only in the highest-$k$ bin.

The analytic mocks are generated with a fixed pixel width of $\Delta v = 10$ km/s, for which the $\sin(x)/x$ term of the window function in Equation \eqref{eq:window_function}, with $x = k\Delta v/2$, significantly suppresses the power at high $k$. 
In Appendix \ref{appendix:high_dv}, we investigate the impact of this effect by repeating the analysis using skewers generated from the same analytic power spectrum with a higher resolution of $\Delta v = 4$ km/s. 
As shown in Figure \ref{fig:appendix_higher_res_compare}, the optimization residuals remain consistent with zero through the highest-$k$ bin considered. 
The FFT residuals also move closer to zero in the highest-$k$ bins, but remain significantly offset from zero at $k \gtrsim 10^{-1}$ s/km.
Further investigation of the FFT measurements is beyond the scope of this work, as the goal is to validate the optimization method for application on data.

\subsection{Simple Forward-Modeled Simulation Mocks}
\label{sec:const_r_const_snr_results}

We next examine the results for the simple forward-modeled mocks described in Section \ref{sec:forward_modeling}. 
Unlike the analytic mocks, these mocks do not have a known input $P(k)$, so we compare the measurements against a reference model constructed from the non-forward-modeled simulation skewers.

For the FFT method, we construct the reference model by applying the same analysis pipeline used for the forward-modeled skewers onto the non-forward-modeled simulation skewers and averaging the resulting power spectra over all 1000 skewers for each $\lammfp$ model. 

For the optimization method, we again use the effective power, $P_{\rm eff}$, defined in Equation \eqref{eq:P_eff}. 
However, calculating $P_{\rm eff}$ requires evaluating the $P(k)$-weighted kernels in Equation \eqref{eq:k_cov_Pk_weight}, and we do not have an analytic expression for the underlying power spectrum. 
Instead, we use the mean FFT power spectrum from the non-forward-modeled simulation skewers, averaged over all 1000 skewers for each $\lammfp$ model. 
We evaluate this power spectrum at the discrete Fourier modes, rather than in the $k$ bins defined in Table \ref{tab:kbins}, and linearly interpolate between these modes to evaluate $P(k)$ when computing the weighted kernels.

At wavenumbers below the fundamental Fourier mode of the simulation box, we hold $P(k)$ constant at its value at the lowest available mode. 
This boundary condition appears reasonable for the $\lammfp = 150$ cMpc model, based on the points shown in Figure \ref{fig:analytic_fit}, although the $\lammfp = 5$ cMpc model shows an increasing trend toward smaller $k$. 
We find that alternative boundary conditions, such as extrapolating the linear interpolation, can change $P_{\rm eff}$ by up to $\sim 5\%$ in the first bin. 
This sensitivity should be kept in mind when interpreting the results, although the adopted prescription is sufficient for validating the optimization method. 
Alternative interpolation methods can also change $P_{\rm eff}$ by a few percent in the highest-$k$ bins, where fewer discrete Fourier modes contribute to each bin, but linear interpolation is adequate for this validation.

As in the analytic mock analysis, we must also specify a fiducial covariance, which depends on a fiducial power spectrum through Equation \eqref{eq:cov_sum}. 
Here, we use the FFT reference model for the corresponding $\lammfp$ model. 
We again verify that variations in the fiducial power spectrum have little impact on the resulting $P_{\rm eff}$, whereas the choice of $P(k)$ used to evaluate the weighted kernels in Equation \eqref{eq:k_cov_Pk_weight} has a substantially larger effect.

The mocks considered in this section each consist of 10 full-length skewers, as described in Section \ref{sec:forward_modeling}, corresponding to an approximate total path length equivalent to that of 30 quasars, each with $\Delta z = 0.1$.

\begin{figure*}
    \centering
	\includegraphics[width=\textwidth]{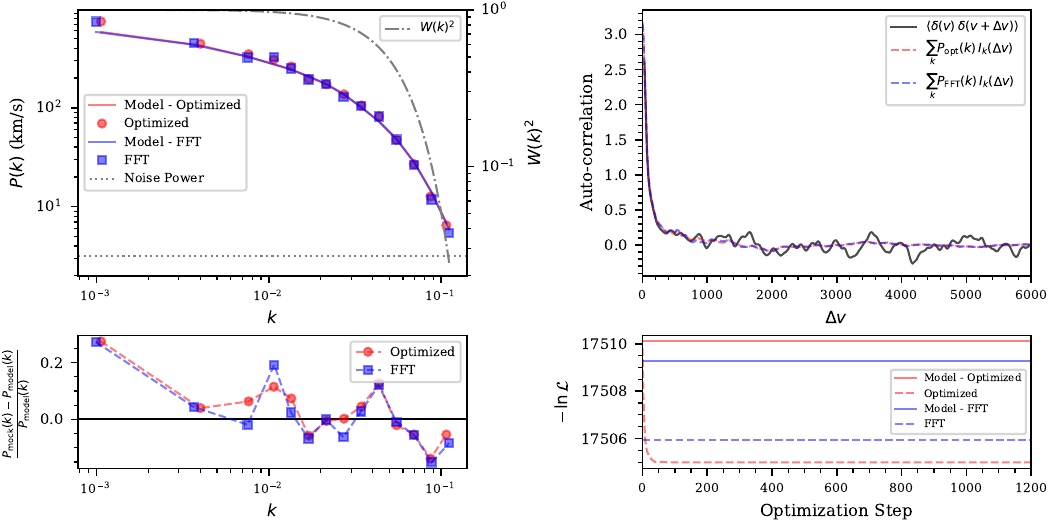}
    \caption{
        Validation on a mock dataset from the $\lammfp = 150$ cMpc simulation comparing FFT and optimization-based measurements of $P(k)$.
        Top left: Measured $P(k)$ from the FFT method corrected for $W_R(k)$ and noise, $P_{\rm FFT}(k)$, (blue squares) and the optimization method, $P_{\rm opt}(k)$, (red circles). 
        Also included are blue and red solid lines representing the model comparison values for the FFT method and the optimization method, respectively. 
        Bottom left: Fractional difference in $P(k)$ relative to the perfect skewers models for the FFT method (blue) and the optimization method (red). 
        Top right: Auto-correlation function from forward-modeled skewers (black), compared to reconstructions from Eq. \eqref{eq:cov_sum} using $P_{\rm opt}(k)$ (red) and $P_{\rm FFT}(k)$ (blue). 
        Bottom right: Minimization of the negative log-likelihood from Equation \eqref{eq:total_like} during optimization (red dashed), compared to the likelihood evaluated at $P_{\rm FFT}(k)$ (blue dashed), and the model value $P_{\rm eff}$ (black solid).
    }
    \label{fig:one_opt}
\end{figure*}

We restrict the analysis to $k \lesssim 10^{-1}$ s/km because, at higher wavenumbers, the noise power becomes comparable to the signal power and the window function falls such that $W_R(k)^2 < 0.04$ (see Figure \ref{fig:one_opt}). 
Consequently, these bins have substantially reduced sensitivity to the underlying signal and are dominated by noise. 
We verified that excluding bins with $k > 10^{-1}$ s/km does not significantly affect the results in the remaining bins, while also reducing computational costs.

An example optimization result is shown in Figure \ref{fig:one_opt} for a single forward-modeled mock dataset from the $\lammfp = 150$ cMpc simulation. 
The top-left panel shows the power spectrum measurements obtained using the FFT method (blue squares) and the optimization method (red circles), while the bottom-left panel shows the corresponding residuals relative to the reference model for each method. 
The grey dotted horizontal line indicates the white-noise power for $\mathrm{SNR}_{\Delta v} = 30$, calculated using Equation \eqref{eq:noise_power}. 
The grey dot-dashed line shows the window function, $W_R(k)^2$, from Equation \eqref{eq:window_function}, plotted against the right-hand axis. 
The residuals from both methods are similar and scatter around zero, remaining within $\sim 25\%$ of their respective model values.

The top-right panel shows the autocorrelation function for this mock dataset. 
The black solid line represents the autocorrelation measured directly from the forward-modeled skewers. 
The red and blue dashed lines show the autocorrelation calculated from the signal covariance matrix in Equation \eqref{eq:cov_sum}, using $P_{\mathrm{opt}}(k)$ and $P_{\mathrm{FFT}}(k)$, respectively. 
The covariance-based predictions capture the overall autocorrelation behavior, particularly at small pixel separations, and reproduce the trends at larger separations while exhibiting less fluctuation than the directly measured autocorrelation.

The bottom-right panel shows the negative log-posterior as a function of optimization step for the likelihood optimization (red dashed line). 
The blue dashed horizontal line indicates the negative log-posterior evaluated using $P_{\mathrm{FFT}}(k)$. 
The red and blue solid horizontal lines show the negative log-posterior evaluated using model values for the optimization and FFT methods, respectively. 
As expected, the optimized solution achieves the lowest negative log-posterior among the cases shown.

\begin{figure*}
    \centering
	\includegraphics[width=\textwidth]{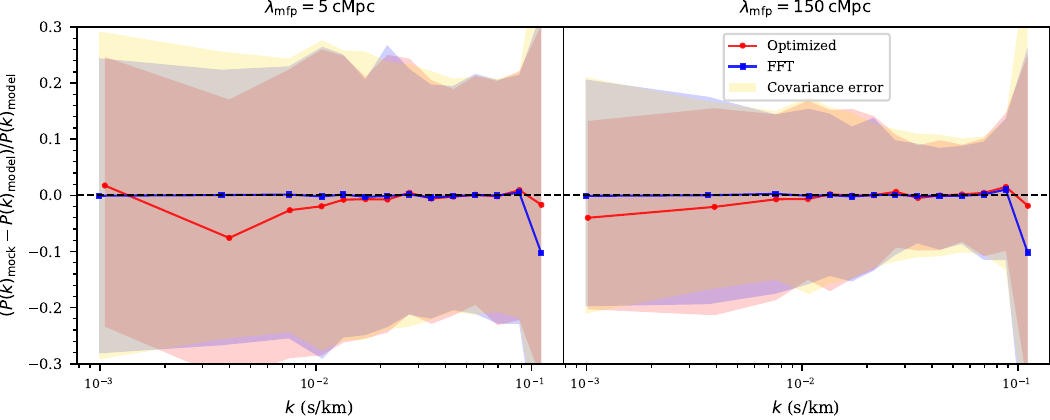}
    \caption{
        The distribution of the difference between the mock $P(k)$ and the model values for both the optimization method (red) and the FFT method (blue). 
        The left column shows mock results from the simulation for $\lammfp = 5$ cMpc while the right column shows the mock results from the simulation for $\lammfp = 150$ cMpc. 
        Both panels show a solid line with the mean value over 100 mock datasets while the shaded region denotes the 16th to 84th percentile representing the $1\sigma$ region. 
        In addition, there is a yellow shaded region showing the expected error from a covariance computation on 500,000 mocks as described by Equation \eqref{eq:power_covariance}.
    }
    \label{fig:sim_opt}
\end{figure*}

While a single optimization realization provides a useful illustration of the estimator, we next examine the distribution of results across 100 independent mock datasets. 
Figure \ref{fig:sim_opt} shows these results for $\lammfp = 5$ cMpc (left column) and $\lammfp = 150$ cMpc (right column). 
We compare the residuals for the FFT method (blue), relative to the reference model constructed from non-forward-modeled simulation skewers, with those for the optimization method (red), relative to the corresponding model $P_{\rm eff}$ values.

To estimate the expected measurement uncertainties for a dataset of this size, we compute the covariance matrix, $\Sigma$, using a large ensemble of mock datasets. 
To reduce computational costs, we use the FFT power spectrum measurements for this calculation. 
The covariance matrix is defined as
\begin{equation}
\Sigma = \frac{1}{N_{\text{mocks}}} \sum_{i=1}^{N_{\text{mocks}}}
\left(P(k)_{\text{FFT},i} - P(k)_{\text{FFT,model}}\right)
\left(P(k)_{\text{FFT},i} - P(k)_{\text{FFT,model}}\right)^{\text{T}},
\label{eq:power_covariance}
\end{equation}
where $N_{\text{mocks}} =$ 500,000 and $P(k)_{\text{FFT,model}}$ is the FFT reference model defined earlier in this section, obtained by averaging the power spectra from 1000 non-forward-modeled simulation skewers. 
The resulting uncertainties are shown by the yellow shaded region centered on 0 in Figure \ref{fig:sim_opt}.

The full residual distributions for both $\lammfp$ models and both measurement methods are consistent with zero and show reasonable agreement with the expected uncertainties estimated from the covariance of 500,000 mocks. 
The mean residuals for the optimization method exhibit small deviations from zero at low $k$, corresponding to large scales.
For $\lammfp = 150$ cMpc, the mean residuals are slightly negative, with a maximum deviation of $\sim 4\%$. 
This is consistent with the expected uncertainty on the mean estimated from the 100 mock datasets and is well within both the $1\sigma$ scatter across mocks and the expected measurement uncertainties. 
The deviation is also comparable to the sensitivity of $P_{\rm eff}$ to the boundary condition used when extrapolating $P(k)$ below the fundamental Fourier mode.
For $\lammfp = 5$ cMpc, the mean residual reaches $\sim 8\%$ in the second $k$ bin, but remains within the expected uncertainty and the $1\sigma$ scatter across mocks. 
This bin contains relatively few discrete Fourier modes, making the result more sensitive to how $P(k)$ is interpolated and integrated within the bin. 
Taken together, these results indicate that the small deviations in the mean residuals are consistent with zero given the uncertainties in the measurements and reference model. 
These considerations will become increasingly important as measurements become more precise.
Model values used when comparing to measurements made with the optimization method must be carefully considered.

\subsection{Masking XQR-30 Data-like Mocks}
\label{sec:xqr30_like_results}

Finally, we investigate the impact of realistic spectral masking on the optimized $P(k)$ measurements. 
We use mock spectra forward-modeled to match the XQR-30 dataset, as described in Section \ref{sec:forward_modeling}, and compare the optimized measurements for each mock with and without the XQR-30 masks applied. 
Unlike the previous tests, this comparison is not intended to assess the accuracy of the measurements relative to a reference model, but rather to isolate the impact of masking on an individual $P_{\rm opt}(k)$ measurement.
Many of the features masked in the XQR-30 data, such as DLAs and quasar proximity regions, are not present in these simulations. 
Therefore, we expect the measurements with and without masking to recover the same underlying power spectrum, with any differences arising from the effects of masking on the measurement.
We also do not investigate the impact of masking on the FFT measurements, as this work focuses on validating the optimization method. 
Examples of the XQR-30 masks and their effects on individual spectra are shown in \cite{wolfson_2023_xqr30}.

\begin{figure}
    \centering
	\includegraphics[width=.5\textwidth]{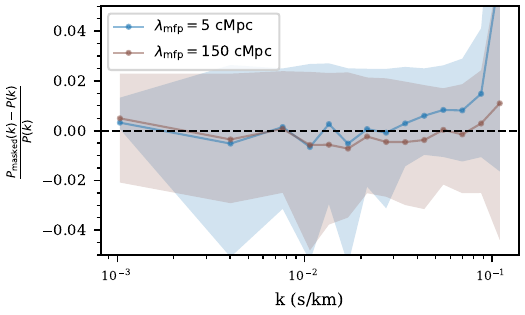}
    \caption{
        The distribution of the differences of $P(k)$ from masked spectra versus unmasked spectra for 100 XQR-30-like mock datasets at $z = 5.7$.
        Blue shows the $\lammfp = 5$ cMpc mock differences while brown shows the $\lammfp = 150$ cMpc model differences.
        The shaded regions show the $1\sigma$ region from the 16th to the 84th percentiles while the solid lines show the means of the distributions. 
    }
    \label{fig:mask_dif}
\end{figure}

The results at $z=5.7$ are shown in Figure \ref{fig:mask_dif}, where the blue and brown curves show the differences for the $\lammfp=5$ and $150$ cMpc models, respectively. 
The solid lines show the mean difference across mock realizations, and the shaded regions show the 16th to 84th percentile range. 
For both models, the shaded regions contain zero across all $k$ bins, indicating no significant systematic difference between the masked and unmasked measurements. 
The typical 16th to 84th percentile range extends from approximately $-0.02$ to $0.02$, suggesting that realistic spectral masking has only a small effect on the recovered $P(k)$ in these mocks.
Although the mean difference for the $\lammfp = 5$ cMpc model trends toward positive values at large $k$ (small scales), the shaded region continues to include zero, and the magnitude of this trend is small compared to the expected cosmic variance in this bin, which is approximately ten times larger.

For the 25 sightlines at $z=5.7$, an average of 10.84 pixels are masked per sightline, corresponding to an average good-pixel fraction of $f_{\rm good}=97.6\%$.
This distribution is highly skewed, however, with most sightlines retaining more than $99\%$ of their pixels and a small number containing substantially larger masked regions.
The expected increase in statistical uncertainty from the reduced number of pixels can be estimated from the good-pixel fraction as
\begin{equation}
\frac{1}{\sqrt{f_{\rm good}}}=1.015.
\end{equation}
Although not shown in the figure, we also calculate the residuals relative to the expected model $P_{\rm eff}$ and measure the width of their 16th to 84th percentile range for the masked and unmasked cases. 
The ratio of the widths of these percentile ranges between the masked and unmasked cases is 1.012 for the $\lammfp=5$ cMpc models and 1.015 for the $\lammfp=150$ cMpc models.
Both are in good agreement with the expected increase in uncertainty from the reduced number of unmasked pixels.
Overall, these results indicate that the realistic XQR-30 masks do not introduce a measurable bias in the recovered $P(k)$, while producing the expected modest increase in statistical uncertainty.

\section{Conclusions}
\label{sec:conclusion}

In this work, we have tested a likelihood-based method for measuring the \lya forest $P(k)$ using three sets of mock observations and compared the results with measurements made using an FFT.

First, in Section \ref{sec:analytic_mock_results}, we considered analytic mock $\delta_F$ skewers generated from two known $P(k)$ models given by Equation \eqref{eq:five_param_power}, with parameters listed in Table \ref{tab:p1d_fit_params}. 
These models were fit to two extreme $\lammfp$ cases from our simulations to represent different levels of large-scale (small-$k$) power. 
Because the input $P(k)$ was known exactly, these mocks provided a direct test of the likelihood optimization against a known underlying model. 
Both methods recovered the input $P(k)$ well, although both had mean residuals with significant deviations from 0 in the highest-$k$ bins. 
Tests using higher-resolution analytic skewers in Appendix \ref{appendix:high_dv} showed that these deviations were resolved for the optimization method. 
The FFT method showed larger deviations across a broader range of high-$k$ bins, but further investigation of the FFT measurements is beyond the scope of this work, which focuses on validating the optimization method. 
Although these Gaussian $\delta_F$ mocks can produce nonphysical values below $\delta_F=-1$, they provide an important validation test by allowing the likelihood method to be assessed against a known input $P(k)$, independently of comparisons between estimators.

Both optimization mean residuals deviated below 0 in the largest $k$ bins but the expected error on the mean, added uncertainty from the $P_{\rm eff}$, and the expected uncertainty on such measurements means that it is not a cause for alarm. 

Next, in Section \ref{sec:const_r_const_snr_results}, we considered simple forward-modeled simulation skewers with fixed resolution, SNR${\Delta v}$, $\Delta v$, and periodic boundary conditions. 
In this case, the true underlying $P(k)$ was not known, so we constructed reference models by measuring the power spectrum of non-forward-modeled simulation skewers using an FFT and then calculating $P{\rm eff}$ from Equation \eqref{eq:P_eff} for comparison with the optimization results. 
For both methods and the $\lammfp$ values considered, the $1\sigma$ regions of the residuals were consistent with zero across the full $k$ range. 
The mean residuals from the optimization method showed some negative deviations in the smallest-$k$ bins. However, these deviations are small compared with the combined uncertainties from the mean of 100 samples and the reference $P_{\rm eff}$ computation, and are consistent with the expected uncertainties in these measurements.

Finally, in Section \ref{sec:xqr30_like_results}, we tested the effect of realistic XQR-30 masking on fully forward-modeled mock spectra.
Comparing the optimized measurements with and without the observational masks showed no significant bias in the recovered $P(k)$.
The masking did increase the scatter across mock datasets, but the increase was consistent with the expected change from the reduced number of unmasked pixels.

One important consideration is that the simulations used here are post-processed with fluctuating UVB fields corresponding to different $\lammfp$ values.
These UVB fluctuations are therefore not coupled to the thermal evolution of the simulations and do not produce the corresponding temperature fluctuations expected from a physically self-consistent reionization history.
As a result, these models may not accurately represent the true $P(k)$ of the high-redshift IGM, particularly near the end of reionization where the thermal memory of reionization can persist.
Nevertheless, they provide a useful test of the likelihood framework by introducing physically motivated large-scale fluctuations in the transmitted flux while allowing us to assess the optimization method across models with different levels of large-scale power.

Overall, we have validated the likelihood-based $P(k)$ measurement method across three increasingly realistic mock-data scenarios where the mock measurements are consistent with the reference model values within expected uncertainties. 
Of note, this likelihood based method accounts for window functions in the transmitted flux field without introducing significant bias into the optimized $P(k)$ measurement.
In future work, we will apply this method to the XQR-30 data to measure the \lya forest $P(k)$ at $z \geq 5$.

\appendix

\section{Optimization Bins}
\label{appendix:bins}

\begin{figure}
    \centering
	\includegraphics[width=.7\textwidth]{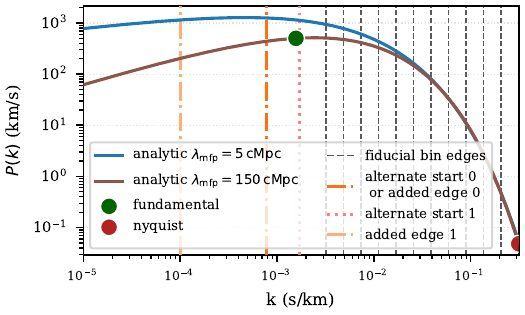}
    \caption{
        The alternate $k$-bin schemes considered in Appendix \ref{appendix:bins}. 
        The solid curves show the analytic model for $P(k)$ from Equation \eqref{eq:five_param_power} for $\lammfp=5$ and $150 \,\mathrm{cMpc}$. 
        The green and red markers indicate the fundamental mode of the mock spectra and the Nyquist frequency, respectively. 
        Gray dashed lines mark the fiducial logarithmic bin edges (the first edge, at $10^{-12},\mathrm{s,km^{-1}}$, is cut off the plot). 
        The colored vertical lines indicate the modified bin schemes. 
        The dark orange dot-dashed line is used in three tests, either as an alternate starting edge or as an additional edge to split up the first fiducial bin. 
        The light orange dot-dashed line is used in the test to split the first fiducial bin into three together with the dark orange line. 
        Finally, the light red dotted line is used as an alternate starting edge that lies above the fundamental mode.
    }
    \label{fig:appendix_bin_edges}
\end{figure}

This appendix investigates how the choice of $k$-bins affect the optimized $P(k)$ measurements. 
We use the analytic mock spectra described in Section \ref{sec:analytic_mocks}, for which the underlying input power spectrum is known exactly, allowing the recovered band powers to be compared directly with the analytic model.

For these tests we adopt a different fiducial $k$-bin scheme than that used in the main text. 
In the main analysis, the bins were chosen to allow for direct comparisons between the likelihood estimator and FFT-based measurements by ensuring that each bin contained at least one discrete Fourier mode. 
Here, however, our goal is simply to compare the optimized band powers with the underlying analytic power spectrum, so considering discrete Fourier modes is unnecessary.
Our fiducial bins therefore consists of a single low-$k$ bin extending from $k=10^{-12}$ s/km to above the fundamental mode of the skewers, followed by 11 logarithmically spaced bins extending to the Nyquist frequency.

The fiducial bin edges are shown by the grey dashed vertical lines in Figure \ref{fig:appendix_bin_edges}, although the first edge at $k = 10^{-12}$ s/km lies outside the plotted range. 
We then consider four classes of alternative bin schemes: 
(i) increasing the minimum $k$ value of the first bin, 
(ii) subdividing the first bin into multiple smaller bins, 
(iii) decreasing the maximum $k$ value by removing the highest-$k$ bins, and 
(iv) changing the number of logarithmically spaced bins between the minimum and maximum $k$ values. 
The first two modifications are illustrated in Figure \ref{fig:appendix_bin_edges}. 
In general, red dotted lines indicate alternate starting edges, while orange dot-dashed lines indicate additional edges used to subdivide the first bin. 
The darker orange dot-dashed line labeled ``alternate start 0 or added edge 0'' corresponds to both an alternate starting edge and a subdivision edge, and is therefore shown only once using the color convention for the additional edges.
The reduced maximum-$k$ schemes are not shown, as they simply remove the final one or two bins.

\begin{figure*}
    \centering
	\includegraphics[width=\textwidth]{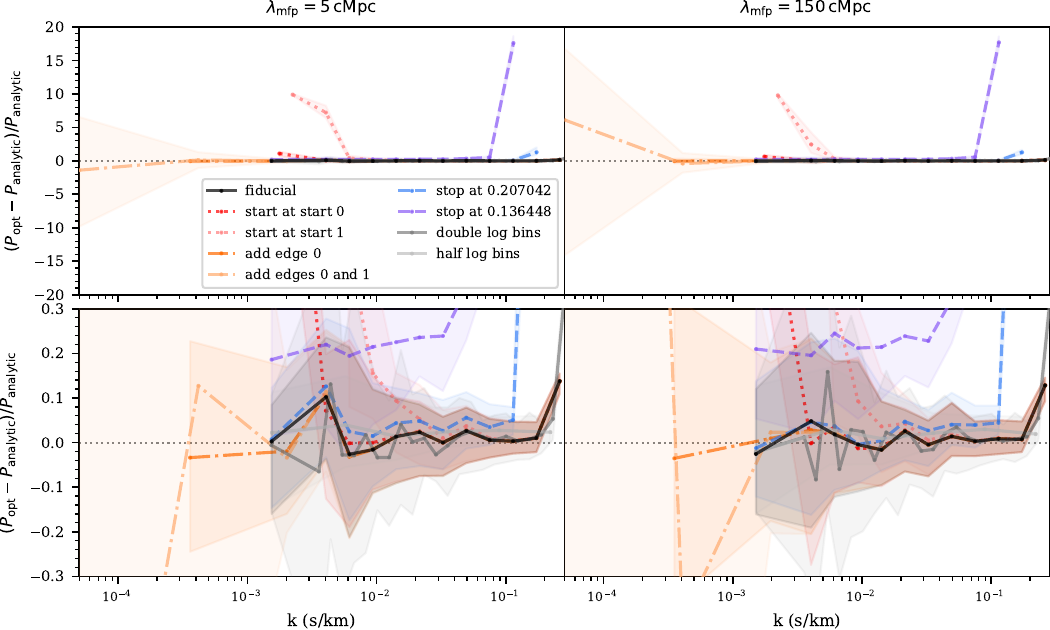}
    \caption{
        Difference between the optimized $P(k)$ using each alternate $k$-bin scheme and the corresponding analytic $P(k)$, normalized by the analytic model, for $\lammfp=5$ (left) and $150\,\mathrm{cMpc}$ (right). 
        The top row shows a $\pm 2000\%$ range which covers the majority of points though it does cut off the top of the purple dashed line and the bottom of the shaded region for the light orange dot dashed line. 
        The bottom row is zoomed to highlight differences within $\pm30\%$.
        The black line shows the fiducial bin results, while the other lines correspond to the alternate bin schemes illustrated in Figure \ref{fig:appendix_bin_edges}. 
        Shaded regions denote the 16th--84th percentile range across 50 mock realizations. 
        Dotted lines correspond to alternate starting bin edges, dot-dashed lines to schemes that subdivide the lowest-$k$ bin, and dashed lines to schemes with a smaller maximum $k$, 
        The gray solid lines correspond to schemes that retain the same first bin while either doubling or halving the number of logarithmically spaced bins over the remaining $k$ range.
        }
    \label{fig:appendix_resid_test}
\end{figure*}

The results for these alternative bin schemes are shown in Figure \ref{fig:appendix_resid_test}. 
For each scheme, we measure $P_{\rm opt}(k)$ for the same 50 independent mock datasets and then compare $P_{\rm opt}(k)$ values with the underlying analytic model as $P_{\rm eff}$ described at the beginning of Section \ref{sec:analytic_mock_results}. 
The resulting scatter therefore reflects both the Gaussian realization of the mock spectra and any additional uncertainty or bias introduced by the optimization procedure and bin choices.
Lines show the mean over the 50 realizations, while the shaded regions indicate the 16th--84th percentile range. 
As in Figure \ref{fig:appendix_bin_edges}, dotted lines correspond to alternate starting bin edges and dot-dashed lines to schemes that subdivide the lowest-$k$ bin. 
In addition, dashed lines correspond to schemes with a reduced maximum $k$ value, while the grey solid lines show the results for schemes that retain the same first bin but use either twice or half as many logarithmically spaced bins thereafter.

The residuals for the fiducial bin scheme are shown by the black line and shaded region. 
The residuals are consistent with zero for bins with $k \lesssim 2 \times 10^{-1}$ km/s, but show a significant positive deviation in the highest-$k$ bin. 
As discussed in Appendix \ref{appendix:high_dv}, tests using $\Delta v = 4$ km/s analytic mocks eliminate this deviation, suggesting that it arises from the finite pixel spacing of the $\Delta v = 10$ km/s mocks. 
For the $\lammfp = 5$ cMpc model, the mean residual in the second bin is also slightly elevated, reaching $\sim 10\%$. 
However, the $1\sigma$ region still contains zero, and the uncertainty on the mean is relatively large because the result is based on only 50 mock realizations. 
Since the goal of this appendix is to compare how different bin choices affect the recovered $P_{\rm opt}(k)$, the fiducial bins provide a baseline for these comparisons. 
As discussed below, most of the alternative bin schemes show a similar positive offset in this bin, suggesting that it is likely a statistical fluctuation rather than an effect specific to the fiducial bins.

We first consider the schemes that subdivide the lowest-$k$ bin. 
Adding only the first additional edge (``edge 0") causes the lowest-$k$ bin to end below the fundamental Fourier mode of the spectra. 
The corresponding results are shown by the darker orange dot-dashed line in Figure \ref{fig:appendix_resid_test}. 
For both $\lammfp$ models, the residual of the lowest-$k$ bin remains consistent with zero, while the $1\sigma$ region is larger than the one in the fiducial model.
A more extreme case is obtained by adding both ``edge 0" and ``edge 1", thereby subdividing the original lowest-$k$ bin into three bins. 
The results are shown by the lighter orange dot-dashed line. 
In this case, the lowest-$k$ bin has a huge amount of scatter making it effectively unconstrained, with a mean residual of approximately $-100\%$ and $1\sigma$ scatter of $\pm600\%$ for $\lammfp = 5$ cMpc, and mean residual of approximately $500\%$ with $1\sigma$ scatter of $\pm1000\%$ for $\lammfp = 150$ cMpc. 
The second bin is also noticeably affected, whereas the third and higher-$k$ bins remain consistent with the fiducial results.
These tests suggest that subdividing the lowest-$k$ bin beyond the scale set by the fundamental Fourier mode produces poorly constrained measurements at the largest scales. 
However, these poorly constrained bins have little impact on the recovered band powers at higher $k$.

We next consider increasing the minimum $k$ value of the first bin. 
The ``start 0'' scheme increases the lower edge of the first bin to half of the fundamental mode of the spectra and is shown by the darker red dotted line. 
Note that this starting edge is shown as an orange dot-dashed line in Figure \ref{fig:appendix_bin_edges} because it also serves as a subdivision edge. 
For this scheme, the mean residual of the first $k$ bin is shifted significantly above zero, with the full $1\sigma$ range remaining above zero. 
The second $k$ bin and all subsequent bins recover behavior consistent with the fiducial bin scheme. 
Increasing the minimum $k$ further with the ``start 1'' scheme, which places the first edge just above the fundamental mode and is shown by the lighter red dotted line, amplifies this effect.
The first several bins show significant positive deviations from both the analytic model and the fiducial bin result shown in black, with agreement only recovered at $k \gtrsim 2 \times 10^{-2}$ s/km.
This test demonstrates the importance of choosing a sufficiently low starting edge for the first bin. 
In particular, the first bin should extend well below the fundamental mode, as even the ``start 0'' case, which begins at half the fundamental mode, produces a biased measurement at the smallest $k$.

When reducing the maximum $k$ value considered, we effectively remove either the final bin (blue dashed line) or the final two bins (purple dashed line).
Removing only the highest-$k$ bin results in agreement with the fiducial bin scheme up to approximately $k \gtrsim 10^{-1}$ s/km, where the final retained bin begins to deviate and reaches a residual of $\sim 200\%$.
Removing the final two bins leads to larger deviations, with the mean residual in the highest retained bin reaching $\sim 20\%$ and the final bin exhibiting extremely large scatter, reaching $\sim 2000\%$.
This demonstrates that excluding the highest-$k$ modes can significantly impact the recovery of the remaining high-$k$ band powers, especially when excluding bins above the level of noise, which is true for the second to last bin as is seen in the top left panel in Figure \ref{fig:analytic_one_opt}.

Finally, we consider changing the number of logarithmically spaced bins while keeping the same initial bin edge.
The darker grey line corresponds to doubling the number of logarithmic bins, while the lighter grey line corresponds to approximately halving the number of bins (reducing from 11 to 5).
Both schemes recover the fiducial bin residual behavior over the full $k$ range.

Overall, these tests motivate the $k$ bins adopted in the main analysis. 
The lowest-$k$ bin must extend below the fundamental mode to avoid biased large-scale measurements, while subdividing this region further produces poorly constrained band powers. 
At high $k$, retaining sufficient bins is important to capture the scale dependence of the power spectrum and avoid averaging over regions where the constant band-power approximation becomes inaccurate.

\section{Higher-resolution Analytic Mocks}
\label{appendix:high_dv}

As discussed in Section \ref{sec:analytic_mock_results}, the residuals between the mock and model $P(k)$ measurements deviate from zero at high $k$. 
The mock skewers are generated on a $\Delta v = 10$ km/s grid, for which the finite pixel width introduces a significant window function correction at high $k$. 
Specifically, the $\sin(x)/x$ term in Equation \eqref{eq:window_function} accounts for the effects of finite spectrograph pixels. 
At the lower edge of the highest-$k$ bin, $k = 0.25$ s/km, this contribution to the window function is
\begin{equation}
\left[\frac{\sin(k\Delta v/2)}{k\Delta v/2}\right]^2
= \left[\frac{\sin(0.25 \times 10/2)}{0.25 \times 10/2}\right]^2
\simeq 0.58,
\end{equation}
representing a substantial suppression of the power. To investigate the impact of this effect, we repeat the analysis using higher-resolution skewers with $\Delta v = 4$ km/s while keeping the total skewer length fixed at $L = 4000$ km/s. 
At the lower edge of the highest-$k$ bin, the window function contribution is now approximately $0.92$, substantially reducing the effect of finite pixel width.

\begin{figure*}
    \centering
	\includegraphics[width=\textwidth]{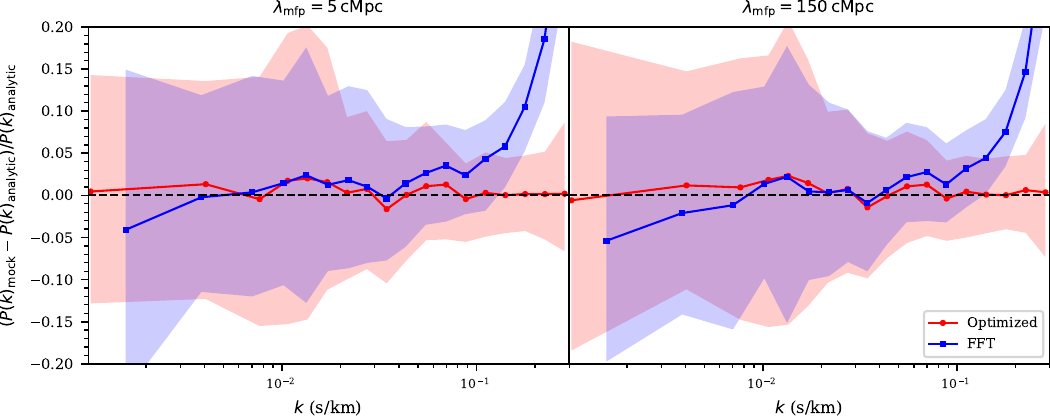}
    \caption{
        Same as Figure \ref{fig:analytic_opt}, but for higher-resolution skewers with $\Delta v = 4$ km/s. 
        The panels show the distribution of residuals between the mock $P(k)$ measurements and the analytic model for the optimization method (red) and FFT (blue), for $\lammfp = 5$ cMpc (left) and $\lammfp = 150$ cMpc (right). 
        Solid lines indicate the mean residual across 100 mock datasets, and shaded regions show the 16th to 84th percentile range. 
        The optimization residuals remain consistent with zero through the highest-$k$ bin considered, while the FFT residuals continue to deviate from zero at high $k$.
    }
    \label{fig:appendix_higher_res_compare}
\end{figure*}

Figure \ref{fig:appendix_higher_res_compare} shows the residuals between the analytic model and the mock power measurements for the higher-resolution skewers, for comparison with the $\Delta v = 10$ km/s results in Figure \ref{fig:analytic_opt}. 
With the higher-resolution skewers, the optimization residuals remain consistent with zero through the highest-$k$ bin considered. This supports the interpretation that the high-$k$ deviation in the optimization residuals in Figure \ref{fig:analytic_opt} arises from neglecting the window function in the analytic model. 
In contrast, the FFT residuals remain offset from zero at high $k$, indicating that the window function alone does not account for the discrepancy in the FFT measurements.

\acknowledgments

MW acknowledges useful conversations with Paul Martini and the CCAPP Cosmology group. 
JFH acknowledges support from the European Research Council (ERC) under the European Union’s Horizon 2020 research and innovation program (grant agreement No 885301).
This research also used resources of the National Energy Research Scientific Computing Center (NERSC), a U.S. Department of Energy Office of Science User Facility located at Lawrence Berkeley National Laboratory, operated under Contract No. DE-AC02-05CH11231.

Generative AI tools (OpenAI’s ChatGPT and Microsoft Copilot) were used to assist with language editing and manuscript revision, the development of plotting scripts used to produce the figures, and the development of the analytic mock function software. 
MW independently validated the analytic mock function software. 
All scientific analyses, results, and conclusions are the authors’ own.



\bibliographystyle{JHEP}
\bibliography{all_papers}

\end{document}